\documentclass[a4paper,fleqn]{cas-sc}

\usepackage[numbers]{natbib}
\hypersetup{colorlinks=true, linkcolor=blue, anchorcolor=blue, citecolor=blue, urlcolor=blue}

\def\D{\,\textrm{d}}
\newcommand{\eref}[1]{Eq.~(\ref{#1})}
\newcommand{\esref}[1]{Eqs.~(\ref{#1})}
\newcommand{\fref}[1]{Fig.~\ref{#1}}
\newcommand{\fsref}[2]{Fig.~\ref{#1}~-~\ref{#2}}
\newcommand{\sref}[1]{Sec.~\ref{#1}}
\newcommand{\tref}[1]{Table~\ref{#1}}
\newcommand{\tsref}[2]{Table~\ref{#1}~-~\ref{#2}}

\begin{document}
\let\WriteBookmarks\relax
\def\floatpagepagefraction{1}
\def\textpagefraction{.001}
\shorttitle{Observational Constraints on Extended Starobinsky and Weyl Gravity Model of Inflation}
\shortauthors{P. Burikham et~al.}

\title[mode = title]{Observational Constraints on Extended Starobinsky and Weyl Gravity Model of Inflation}

\author[1]{{\color{black}Piyabut} Burikham}[type=editor,
                        auid=000,
                        bioid=1,
                        role=]
\ead{piyabut@gmail.com}

\credit{Conceptualization of this study}

\author[2]{{\color{black}Teeraparb} Chantavat}[type=editor, orcid=0000-0002-0259-1591]
\cormark[1]
\ead{teeraparbc@nu.ac.th}

\credit{Data analysis, Writing - Original draft preparation}

\author[1,3]{{\color{black}Pongsapat} Boonaom}[%
   role=,
   suffix=,
   ]
\ead{j.pongsapatb@gmail.com}

\credit{Data analysis}

\affiliation[1]{organization={High Energy Physics Theory Group, Department of Physics, Faculty of Science, Chulalongkorn University},
                city={Bangkok},
                postcode={10330}, 
                country={Thailand}}

\affiliation[2]{organization={The Institute for Fundamental Study, Naresuan University},
                city={Phitsanulok},
                postcode={65000}, 
                country={Thailand}}

\affiliation[3]{organization={National Astronomical Research Institute of Thailand (NARIT)},
                city={Chiang Mai},
                postcode={50200}, 
                country={Thailand}}

\cortext[cor1]{Corresponding author}

\begin{abstract}
We present constraints on the extended Starobinsky and Weyl gravity model of inflation using updated available observational data. The data includes cosmic microwave background (CMB) anisotropy measurements from {\it Planck} and BICEP/{\it Keck} 2018 (BK18), as well as large-scale structure data encompassing cosmic shear and galaxy autocorrelation and cross-correlation functions measurements from Dark Energy Survey (DES), baryonic acoustic oscillation (BAO) measurements from 6dF, MGS and BOSS, and  distance measurements from supernovae type Ia from Pantheon+ samples. By introducing a single additional parameter, each model extends the Starobinsky model to encompass larger region of parameter space while remaining consistent with all observational data.  Our findings demonstrate that the inclusion of higher-order terms loosen the constraint on the upper bound of $e$-folding number $N_{\rm e}$ due to the presence of small additional parameter.  The maximum limit on $N_{\rm e}$ could be refined by considering the reheating process to $N_{\rm e}<55-59$ for $k_{*}=0.002, 0.05$ Mpc$^{-1}$.  These models extend viable range of tensor-to-scalar ratio~($r$) to very small value $r<0.002$ in contrast to the original $R^2$ Starobinsky model. In addition, our results continue to emphasize the tension in $H_0$ and $S_8$ between early-time CMB measurements and late-time large-scale structure observations.
\end{abstract}


\maketitle

\section{Introduction}

The Starobinsky model is one of the simplest models of inflation that is consistent with observational constraints from Planck Cosmic Microwave Background radiation 2018~\citep{Planck2018_10}.  The model contains one-loop quantum gravity correction which naturally leads to an $R^{2}$ term in the effective gravity action, in addition to the Einstein-Hilbert linear action term~\citep{Starobinsky1980, Bamba_ea2014, Antoniadis_ea2015}.  The loop contribution is expected to be significant in the very early universe since Planckian physics inevitably introduce higher order terms in the form of $R^{n}$, amongst the other possible Lorentz invariant combinations including the form of $f(R)$~\cite{Myrzakulov_ea2015, Myrzakulov_ea2016, Odintsov_ea2017, Odintsov_Oikonomou2018, Odintsov_Oikonomou2019a, Odintsov_Oikonomou2019b, Elizalde_ea2019, Nojiri_ea2017}, into the effective gravity action~(see \cite{Burgess_2004} for a nice review and \cite{Gialamas_Lahanas2020,Gialamas_ea2020,Gialamas_ea2021,Gialamas_ea2023} for inflationary modes in Palatini formalism).  These higher order terms should play an important role in very early stage of the universe, potentially causing and governing the inflationary era. 

A phenomenological attempt to consider the effects of $R^{3}$ term is proposed in \cite{Cheong_ea2020} where the coefficient of the extra term is taken to be a free small parameter with respect to the $R^{2}$ contribution and the prospect of observational constraints is studied in Ref.~\cite{Modak_ea2023}. The model is an extension of the Starobinsky model where unitary is not violated as long as the $R^{3}$ term is kept relatively small comparing to the dominating $R^{2}$ term. In contrast, a conformal Weyl inflation model is proposed in \cite{Wang_ea2023} with different structure of the $R^{2}$ and $R^{3}$ terms. Both models can be set to reduce to conventional Starobinsky model but their extensions to the $R^{3}$ contribution are different and deserve detailed comparison with respect to the updated observational data. In this work, we consider constraints from Planck CMB 2018 (TTTEEE+lowE+lensing), BK18 (Bicep Keck 2018)~\cite{Ade_ea2021}, large-scale structure data, i.e., BAO (Baryon Acoustic Oscillation)~\cite{Alam_ea2017}, DES (Dark Energy Servey)~\cite{Abbott_ea2018}, and low-redshift Pantheon+ supernovae type Ia sample \cite{Scolnic_ea2022} to the parameter space of Starobinsky, $R^3$ extended Starobinsky, and Weyl gravity model of inflation.

This work is organized as follows: In \sref{sec:r2r3models}, we provide a review of the basic scalar-tensor gravity theory and extended Starobinsky model. Subsequently, we compute the inflationary model parameters. In \sref{sec:Wmodel}, we introduce Weyl gravity model of inflation along with the relevant inflationary parameters.  We explain the data analysis and provide a concise overview of the data we employed in \sref{sec:dataanalysis}.  The results are presented in \sref{sec:results}.  We discuss our results and provide a concluding summary in \sref{sec:discussion}.

\section{The scalar-tensor gravity theory and Starobinsky models}
\label{sec:r2r3models}

We start with an overview of the well-known results in scalar-tensor gravity theory~\cite{Sebastiani_ea2014,Canko_ea2020}.  In the Jordan frame, the action of matter and generic $f(R)$ gravity can be expressed as 
\begin{eqnarray}
    \nonumber S &=& S_{\rm J} + S_{m}(g_{{\rm J}\mu\nu},\Psi_{m}), \\
    &=& \frac{1}{2\kappa^2}\int{\D^{4}x \sqrt{-g_{\rm J}} f(R_{\rm J})} +  S_{m}(g_{{\rm J}\mu\nu},\Psi_{m}),
\end{eqnarray}
where $\kappa^2 = 8\pi G = 1$, $S_{m}$ is a matter action with fields $\Psi_{m}$ and $g_{\rm J}$ is the determinant of the spacetime metric $g_{\rm J\mu\nu}$. Performing the Legendre transformation to the gravity part of the action gives rise to
\begin{equation}
    S_{\rm J} = \frac{1}{2}\int{\D^{4}x \sqrt{-g_{\rm J}}[F(R_{\rm J})R_{\rm J}-U(R_{\rm J}))]},
\end{equation}
where $F(R_{\rm J}) \equiv \partial f/\partial R_{\rm J}$ and $U(R_{\rm J}) = F(R_{\rm J})R_{\rm J}-f(R_{\rm J})$. 

The $f(R)$ gravity is equivalent to the scalar-tensor theory by a conformal transformation
\begin{equation}
g_{\rm E\mu\nu}=\Omega^2g_{\rm J\mu\nu} \rightarrow \sqrt{-g_{\rm E}}=\Omega^4\sqrt{-g_{\rm J}},
\end{equation}
where $\Omega^2$ is a conformal factor.
Under the transformation, we obtain the transformed Ricci scalar 
\begin{eqnarray}
    R_{\rm J}&=& \Omega^2R_{\rm E} + 6g_{\rm E}^{\mu\nu}\Omega(\partial_{\mu}\partial_{\nu}\Omega) -12g_{\rm E}^{\mu\nu}(\partial_{\mu}\Omega)(\partial_{\nu}\Omega), \notag \\
    &\equiv& \Omega^2[R_{\rm E} - g_{\rm E}^{\mu\nu}(\partial_{\mu}s)(\partial_{\nu}s)],
\end{eqnarray}
where $s$ is a canonical field.  The gravity action in the Einstein frame then takes the form
\begin{equation}
S_{\rm E}=\int{\D^{4}x \sqrt{-g_{\rm E}}\bigg[\frac{1}{2}R_{\rm E}-\frac{1}{2}g_{\rm E}^{\mu\nu}\partial_\mu s \partial_\nu s - V_{\rm E}(s) \bigg]},
\end{equation}
where we choose
\begin{equation}
    \Omega^2(\phi) = F(R_{\rm J}) = f'(R_{\rm J}) = f'(\phi), 
\end{equation}
and a new scalar field $R_{\rm J} \equiv \phi$ is defined. The canonical field $s(\phi)$ can then be expressed as 
\begin{equation}
    s(\phi) = \sqrt{\frac{3}{2}} \ln{\Omega^2(\phi)}=\sqrt{\frac{3}{2}} \ln{[f'(\phi)]}. \label{seqn}
\end{equation}
And the potential in the Einstein frame is
\begin{eqnarray}
    \nonumber V_{\rm E}(s) = \frac{U(R_{\rm J})}{2\Omega^4} &=& \frac{F(R_{\rm J})R_{\rm J}-f(R_{\rm J})}{2\Omega^4}, \\
    &=& \frac{\phi f'(\phi)-f(\phi)}{2f'(\phi)^2} \Big|_{\phi=\phi(s)}.  \label{VEeqn}
\end{eqnarray}

\subsection{The extended Starobinsky model}
\label{ssec:extendedStar}

First, we review the extended Starobinsky model studied in \cite{Cheong_ea2020} where a slightly different approximation with respect to the $e$-folding number $N_{\rm e}$ is used in our calculation in \sref{ssec:ESmodel}. Start with
\begin{equation}
    f(R)=aR+bR^2+cR^3,
\end{equation}
where $a=1$, $b=\displaystyle{\frac{\beta}{2}}$ and $c=\displaystyle{\frac{\gamma}{3}}$. The conformal factor becomes 
\begin{equation}
\Omega^2(\phi) = f'(\phi) = 1+\beta \phi+\gamma \phi^2.
\end{equation}
Substitute into \eref{seqn} to obtain $s(\phi)$
\begin{equation}
    s(\phi)=\sqrt{\frac{3}{2}} \ln{[1+\beta \phi + \gamma \phi^2]}, \notag
\end{equation}
and define the quadratic field of scalar
\begin{equation}
    \sigma(s) \equiv \exp\left(\sqrt{\frac{2}{3}}s\right) = 1+\beta \phi +\gamma \phi^2.
\end{equation}
The $\phi(s)$ can be solved as a solution of the quadratic equation
\begin{equation}
    \phi(s) = \frac{\beta}{2\gamma} \left( \sqrt{1+4\frac{\gamma}{\beta^2}(\sigma(s)-1)} -1 \right).
\end{equation}
Consider if $\gamma \ll \beta$ and $\phi \approx 1$, we impose $\gamma$ as a small perturbation in $\sigma(s)$, so that
\begin{displaymath}
    \beta\phi(s) + 1 = \sigma(s) -\frac{\gamma}{\beta^2}(\sigma(s)-1)^2 + \cdots.
\end{displaymath}
Solve the equation above to obtain $\phi(s)$
\begin{equation}
\phi(s) = \frac{\sigma(s) -1}{\beta} \left[ 1-\frac{\gamma}{\beta} \left( \frac{\sigma(s)-1}{\beta} \right) + \mathcal{O} \left( \frac{\gamma}{\beta} \right) ^2 \right].
\end{equation}
The potential in Einstein frame \eref{VEeqn} then becomes
\begin{eqnarray}
V_{\rm E}(s) &&= \frac{\beta\phi(s)^2(1+\frac{4\gamma}{3\beta}\phi(s))}{4(1+\beta\phi(s)(1+\frac{\gamma}{\beta}\phi(s)))^2}, \notag \\
&&\approx V_0(s) \left[ 1-\frac{2\gamma}{3\beta} \left( \frac{\sigma(s) -1}{\beta} \right) + \cdots \right], 
\end{eqnarray}
where $V_0(s) = \displaystyle{\frac{1}{4\beta}\left(1-\frac{1}{\sigma}\right)^2 = \frac{1}{4\beta}\left(1-e^{-\sqrt{\frac{2}{3}}s}\right)^2}$ for $\gamma = 0$ .

\subsection{The slow-roll inflation}
\label{ssec:ESmodel}

The slow-roll parameters can be approximate to the leading order of $\mathcal{O}(\gamma/\beta)$ as
\begin{eqnarray}
    \epsilon &=& \frac{1}{2} \left( \frac{V'_{\rm E}}{V_{\rm E}} \right)^2 = \epsilon_0 + \frac{\gamma}{\beta} \Delta\epsilon,  \\ 
    \eta &=& \frac{V''_{\rm E}}{V_{\rm E}} = \eta_0 + \frac{\gamma}{\beta} \Delta\eta,
\end{eqnarray}
where $\epsilon_0$ and $\eta_0$ are the slow-roll parameters for $\gamma = 0$ and we calculate them in terms of perturbation
\begin{eqnarray}
    \epsilon_0 &=& \frac{4}{3(\sigma(s)-1)^2},\\
    \eta_0 &=& -\frac{4(\sigma(s)-2)}{3(\sigma(s)-1)^2},\\
    \Delta\epsilon &=& -\frac{8\sigma(s)}{9\beta(\sigma(s)-1)} + \mathcal{O} \left( \frac{\gamma}{\beta} \right),\\
    \Delta\eta &=& -\frac{4\sigma(s)(\sigma(s)+3)}{9\beta(\sigma(s)-1)} + \mathcal{O} \left( \frac{\gamma}{\beta} \right).
\end{eqnarray}
The number of $e$-foldings from the start to the end of inflation $(s_{\rm e} < s)$ can then be determined
\begin{eqnarray}
N_{\rm e} (s) &=& \int^{s}_{s_e} \frac{\D s}{\sqrt{2\epsilon}} = N_{\rm e,0} + \Delta N_{\rm e},  \notag \\
&\approx& \frac{3}{4} \sigma(s)+\left( \frac{\gamma}{\beta} \right) \frac{\sigma(s)^3}{12\beta}
\end{eqnarray}
where $N_{\rm e,0}$ is $e$-folding number for $\gamma = 0$ and $\Delta N_{\rm e}$ is correction at the leading order $\gamma/\beta$ and assuming $s \gg s_e \sim 1$.  For $N_{\rm e}$ $e$-foldings at $s_{*}$,
\begin{equation}
N_{\rm e}(s_*) = \frac{3}{4} \sigma(s_*) + \left( \frac{\gamma}{\beta} \right) \frac{\sigma(s_*)^3}{12\beta}.
\end{equation}
The asymptotic solution of $\sigma(s_*)$ for generic $N_{\rm e}$ is thus
\begin{equation}
    \sigma(s_*)\equiv \sigma_{*} \approx \frac{4}{3} N_{\rm e} - \delta \frac{64}{243} N_{\rm e}^{3},
\end{equation}
where $\delta \equiv \gamma/\beta^2 \ll 1$. The inflaton vacuum energy at horizon exit can be estimated from COBE normalization to be~\cite{Cheong_ea2020}
\begin{equation}
\frac{V_{*}}{\epsilon(s_{*})}\approx \frac{\sigma_{*}^{2}}{\beta}\left(\frac{3}{16} + \delta \frac{\sigma_{*}^{2}}{8}\right) = 24\pi^{2}A_{\rm s}M^{4}_{\rm P} \simeq 0.027^{4} M_{\rm P}^{4},    \label{Vh:eq}
\end{equation}
where $\ln(10^{10}A_{\rm s})\simeq 3.10$~\cite{Planck2018_10} is imposed.
We can thus solve to obtain
\begin{equation}\label{be:eq}
\beta \simeq \frac{N_{\rm e}^{2}}{0.027^{4} M_{\rm P}^{4}}\left( \frac{1}{3}+\frac{64}{243}\delta N_{\rm e}^{2}\right).    
\end{equation}

The primordial power spectra (scalar and tensor mode) are parameterized in power-law forms as follows: 
\begin{equation}
    \label{eq:power_scalar}
    \ln\mathcal{P}_{\rm s}(k) = \ln A_{\rm s} + (n_{\rm s} - 1) \ln\left(\frac{k}{k_*}\right) + \frac{1}{2} \frac{{\rm d}\ln n_{\rm s}}{{\rm d}\ln k} \ln \left(\frac{k}{k_*} \right)^2 + \frac{1}{6} \frac{{\rm d}^2\ln n_{\rm s}}{{\rm d}\ln k^2} \ln\left(\frac{k}{k_*} \right)^3 + \ldots,
\end{equation}
and
\begin{equation}
    \label{eq:power_tensor}
    \ln\mathcal{P}_{\rm t}(k) = \ln(r A_{\rm s}) + n_{\rm t} \ln\left(\frac{k}{k_*} \right) + + \frac{1}{2} \frac{{\rm d}\ln n_{\rm t}}{{\rm d}\ln k} \ln \left(\frac{k}{k_*} \right)^2 + \ldots,
\end{equation}
where $\mathcal{P}_{\rm s}$ and $\mathcal{P}_{\rm t}$ are the scalar and tensor power spectrum respectively.  $n_{\rm s}$ and $n_{\rm t}$ are the scalar and tensor spectral index.  $r$ is the tensor-to-scalar ratio and $A_{\rm s}$ is the primordial scalar power spectrum amplitude.  We define the running and running of running of the corresponding parameters as $\displaystyle n_{\rm run} \equiv {\rm d}\ln n_{\rm 
s}/{\rm d}\ln k$, $\displaystyle n_{\rm run, run} \equiv {\rm d}^2\ln n_{\rm s}/{\rm d}\ln k^2$ and $n_{\rm t, run} \equiv {\rm d}\ln n_{\rm t}/{\rm d}\ln k$.

The observable cosmological parameters are obtained in terms of two free parameters $N_{\rm e}$ and $\delta$. 
The scalar spectral index is given explicitly by
\begin{eqnarray}
    \label{eq:ns_star}
    n_{\rm s} &&= 1-6\epsilon(s_*) +2\eta(s_*), \notag  \\ 
    &&=\frac{16N_{\rm e}^2 - 56N_{\rm e} - 15}{(4N_{\rm e} - 3)^2}  - \delta \frac{32(256N_{\rm e}^4 - 576N_{\rm e}^3 + 756N_{\rm e}^2 - 243N_{\rm e})}{81(4N_{\rm e} - 3)^3}. \\
    \notag 
\end{eqnarray}
Note that we keep the full dependence on $N_{\rm e}$. The tensor to scalar ratio is thus
\begin{eqnarray}
    \label{eq:r_star}
    r &&= 16\epsilon(s_*), \notag \\
    \nonumber && = \frac{192}{(4N_{\rm e} - 3)^2} - \delta \frac{512(32N_{\rm e}^3 - 72N_{\rm e}^2 + 27N_{\rm e})}{27(4N_{\rm e} - 3)^3}. \\
\end{eqnarray}
Moreover the running of scalar spectral index is
\begin{eqnarray}
    \label{eq:nrun_star}
    n_{\rm run} &&=\alpha_s = \frac{\D n_{\rm s}}{\D \log k} = -2\xi +16\epsilon\eta -24\epsilon^2, \notag \\
    &&= -\frac{128(4N_{\rm e}^2 + 9N_{\rm e}) }{(4N_{\rm e} - 3)^4} + \delta \frac{128(1024N_{\rm e}^5 - 3840N_{\rm e}^4 + 864N_{\rm e}^3 - 1620N_{\rm e}^2 +729N_{\rm e})}{81(4N_{\rm e} - 3)^5},
\end{eqnarray}
and the running of running of scalar spectral index is given by
\begin{eqnarray}
    \label{eq:nrunrun_star}
    n_{\rm run,run}=\quad\beta_s &&= \frac{d \alpha_s}{d \log k} =  -2\omega -2\eta\xi +24\epsilon\xi +32\epsilon\eta^2 -192\epsilon^2\eta +192\epsilon^3, \notag \\
    &&= \frac{512(32N_{\rm e}^3 + 132N_{\rm e}^2 + 27N_{\rm e})}{(4N_{\rm e} - 3)^6} \notag \\
	&&+ \delta \frac{512(2048N_{\rm e}^5 + 20352N_{\rm e}^4 + 864N_{\rm e}^3 - 1620N_{\rm e}^2 - 729N_{\rm e})}{27(4N_{\rm e} - 3)^7},
\end{eqnarray}
where~\cite{Liddle_Lyth2000}
\begin{eqnarray}
\xi &\equiv& \displaystyle{\frac{V'_{\rm E} V'''_{\rm E}}{V_{\rm E}^2}}=\frac{16 (\sigma -4)}{9 (\sigma -1)^3}-\delta \frac{16 \sigma  \left(2 \sigma ^2-7 \sigma +11\right)}{27 (\sigma -1)^3}, \notag \\
\omega &\equiv& \displaystyle{\frac{V'_{\rm E}{}^2V''''_{\rm E}}{V_{\rm E}^3}}=-\frac{64 (\sigma -8)}{27 (\sigma -1)^4}+\delta  \frac{64 \sigma  \left(\sigma ^2-20 \sigma +31\right)}{81 (\sigma -1)^4}. \notag
\end{eqnarray}
The tensor spectral index can also be computed,
\begin{eqnarray}
    \label{eq:nt_star}
    n_{\rm t} &&= -2\epsilon \notag \\
    \nonumber &&= - \frac{24}{(4N_{\rm e}-3)^2} +  \delta \frac{64(32N_{\rm e}^3 - 72N_{\rm e}^2 + 27N_{\rm e})}{27(4N_{\rm e} - 3)^3}. \\
\end{eqnarray}
And the running of tensor spectral index takes the form,
\begin{eqnarray}
    \label{eq:ntrun_star}
    n_{\rm t,run} &&= 4\epsilon\eta - 8\epsilon^2 \notag \\
    &&=-\frac{768 N_{\rm e}}{\left(4 N_{\rm e} - 3\right)^4}\notag \\
    && -\delta \frac{256 \left(64 N_{\rm e}^3+36 N_{\rm e}^2-27 N_{\rm e}\right)}{9 \left(4 N_{\rm e}-3\right){}^5}.
\end{eqnarray}
For $\delta = 0$, the Starobinsky model $aR+bR^2$ is recovered,
\begin{eqnarray}
    \label{eq:ns_Ne}
    n_{\rm s} &&=\frac{16N_{\rm e}^2 - 56N_{\rm e} - 15}{(4N_{\rm e} - 3)^2} \simeq 1 - \frac{2}{N_{\rm e}} - \frac{9}{2N_{\rm e}^2},\\
    \label{eq:r_Ne}
    r &&= \frac{192}{(4N_{\rm e} - 3)^2} \simeq \frac{12}{N_{\rm e}^2}.
\end{eqnarray}

\section{Weyl Gravity Model}
\label{sec:Wmodel}

In comparison to the extended Starobinsky model discussed above, there is another type of extension which contains $R^{2}$ and $R^{3}$ terms, the Weyl gravity~(WG) model. WG model contains additional Weyl scalar and vector fields whence transition to Einstein gravity is achieved after conformal symmetry breaking. On the galactic scale, Weyl (geometric) gravity models~\citep{Ghilencea2019a, Ghilencea_Lee2019, Ghilencea2019b, Ghilencea2020a, Ghilencea2020b, Ghilencea2021, Ghilencea_Harko2021, Ghilencea2022, Ghilencea2023, Weisswange_ea2023} provides alternative possibility of/to dark matter as a successful quantitative description of the galaxy rotation curves~\citep{Burikham_ea2023}.

Details of the inflationary scenario of a class of WG model are explored in \cite{Wang_ea2023} where two inflation scenarios are considered, inflation to the side and inflation to the centre. Here we consider only the inflation to the side scenario with $\zeta \to \infty$. 

For Weyl gravity action, the transformation to Einstein frame of the $f(R)$ gravity in \sref{sec:r2r3models} can be generically performed with replacement 
\begin{equation}
\frac{f(R_{\rm J})}{\kappa^{2}}\to F(R_{\rm J},\varphi)-\zeta D^{\mu}\varphi D_{\mu}\varphi-\frac{1}{2g_{\rm W}^{2}}F_{\mu\nu}F^{\mu\nu},
\end{equation}
where $F_{\mu\nu}=\partial_{\mu}W_{\nu}-\partial_{\nu}W_{\mu}$ for 
\begin{eqnarray}
F(R_{\rm J},\varphi)&=&\varphi^{2}R_{\rm J}+\alpha R_{\rm J}^{2}+\frac{\beta}{\varphi^{2}}R_{\rm J}^{3}. \\ \notag
\end{eqnarray}
This leads to~\cite{Tang_Wu2020, Wang_ea2023} 
\begin{eqnarray}
\frac{\mathcal{L}}{\sqrt{-g_{\rm E}}}&=&\frac{R}{2}-V_{\rm E}(\varphi)-\frac{\partial_{\mu}\varphi \partial^{\mu}\varphi}{2/\zeta+\varphi^{2}/3} - \frac{1}{4g_{\rm W}^{2}}F_{\mu\nu}F^{\mu\nu}  \notag \\
&&+h(\varphi,\zeta,W_{\mu}),
\end{eqnarray}
where $h(\varphi,\zeta,W_{\mu})$ and $V_{\rm E}(\varphi)$ are given in Eq.~(8) and (9) of \cite{Wang_ea2023} respectively.

The potential in the Einstein frame in $\zeta \to \infty$ limit is then given by
\begin{eqnarray}
V_{\rm E}(\Phi)&=&\frac{1}{8 \alpha}\left(1-e^{\sqrt{\frac{2}{3}} \left(\Phi -\Phi _0\right)}\right)^2 \label{VWeq} \\
&&\times\left(1+\frac{1}{6} \left(1-e^{-\sqrt{\frac{2}{3}} \left(\Phi -\Phi _0\right)}\right) \gamma _{\rm W}\right)+\mathcal{O}(\gamma_{\rm W}^{2}),\notag 
\end{eqnarray}
where $\gamma_{\rm W}\equiv \displaystyle{\frac{3\beta}{\alpha^{2}}}$ and the scalar field is redefined by
\begin{equation}
\varphi^{2}\equiv\frac{6}{|\zeta|}\sinh^{2}\left( \frac{\pm \Phi}{\sqrt{6}}\right).
\end{equation}
We can approximate
\begin{equation}
    \Phi -\Phi_{0} \approx - \sqrt{\frac{3}{2}} \ln\Bigg| \sqrt{\frac{12}{\gamma_{\rm W}}\tanh\Big( \tanh^{-1}(0.622 \gamma_{\rm W})+\sqrt{\frac{4\gamma_{\rm W}}{27}}(N_{\rm e}+2.7})\Big) \Bigg| \equiv \sqrt{\frac{3}{2}}\ln \Theta(\gamma_{\rm W},N_{\rm e})
\end{equation}
 for $N_{\rm e}=50-80$ with a slight change of the constant $2.7$ which do not affect the approximation significantly. The cosmological parameters in this model can be straightforwardly computed,
\begin{eqnarray}
    \label{eq:weyl_ns}
    n_{\rm s}&=&\frac{36 (\Theta -5) (3 \Theta +1)+(\Theta -1)^2 ((\Theta -26) \Theta -5) \gamma_{\rm W}^2-60 [\Theta  ((\Theta -5) \Theta +3)+1] \gamma_{\rm W}}{3 (\Theta -1)^2 \left[(\Theta -1) \gamma_{\rm W}-6\right]^2}, \label{nsWeyl} \\
    r &=&\frac{16 \left[\left(\Theta ^2 +\Theta -2\right) \gamma_{\rm W} -12\right]^2}{3 (\Theta -1)^2 \left[(\Theta -1) \gamma_{\rm W} -6\right)]^2}, \label{rWeyl} \\
    n_{\rm run}&=&-\frac{8 \Theta  \left(\left(\Theta ^2+\Theta -2\right) \gamma_{\rm W} - 12\right)}{3 (\Theta -1)^4 \left[(\Theta -1) \gamma_{\rm W} -6\right]^4} \notag \\
    &&\times\Big[(\Theta -1) \gamma_{\rm W} \left(12 (\Theta  (\Theta +4)+21)+(\Theta -1) \gamma_{\rm W} \left(4 (\Theta -6) (\Theta +2)+\left(2 \Theta ^2+\Theta -3\right) \gamma_{\rm W}\right)\right)-144 (\Theta +3)\Big], \notag \\
    \\
    n_{\rm run,run}&=&\frac{32 \Theta  \left(\left(\Theta ^2+\Theta -2\right) \gamma_{\rm W} -12\right)}{9 (\Theta -1)^6 \left[(\Theta -1) \gamma_{\rm W} -6\right]^6} \notag \\
    &&\times \Bigg[\left(-10368 \Theta ^2-57024 \Theta -15552\right)+\left(11664 \Theta ^3+39312 \Theta ^2-36720 \Theta -14256\right) \gamma_{\rm W}  \notag \\
    &&+\left(72 \Theta ^6-792 \Theta ^5-2520 \Theta ^4-12312 \Theta ^3+29808 \Theta ^2-9072 \Theta -5184\right) \gamma_{\rm W}^2 \notag \\
    &&+\left(48 \Theta ^7-180 \Theta ^6+1128 \Theta ^5+1080 \Theta ^4-7992 \Theta ^3+7908 \Theta ^2-1056 \Theta -936\right) \gamma_{\rm W}^3 \notag \\
    &&+\left(2 \Theta ^8+10 \Theta ^7-204 \Theta ^6+284 \Theta ^5+550 \Theta ^4-1422 \Theta ^3+920 \Theta ^2-56 \Theta -84\right) \gamma_{\rm W}^4 \notag \\
    &&+\left(\Theta ^8+5 \Theta^7-24 \Theta ^6+13 \Theta ^5+50 \Theta ^4-81 \Theta^3+40 \Theta ^2-\Theta -3\right) \gamma_{\rm W}^5 \Bigg], \\
    n_{\rm t}&=&-\frac{2 \left[\left(\Theta ^2+\Theta -2\right) \gamma_{\rm W}-12\right]^2}{3 (\Theta -1)^2 \left[(\Theta -1) \gamma_{\rm W}-6\right]^2}, \\
    \label{eq:weyl_ntrun}
    n_{\rm t, run}&=&-\frac{8 \Theta  \left(\left(\Theta ^2+\Theta -2\right) \gamma_{\rm W}-12\right){}^2 \left[2 \left(\Theta ^2-6 \Theta +5\right) \gamma_{\rm W} + (\Theta -1)^2 \gamma_{\rm W}^2+24\right]}{3 (\Theta -1)^4 \left[(\Theta -1) \gamma_{\rm W}-6\right]^4}.  
\end{eqnarray}

The cosmological parameters given in \sref{sec:r2r3models} and \sref{sec:Wmodel} are then subject to observational constraints as discussed in \sref{sec:results}.

\begin{table*}\centering
\caption{\label{tab:dataset} The datasets employed in our work.}
\begin{tabular}{lc}
    \noalign{\vskip 3pt}\hline\noalign{\vskip 1.5pt}\hline\noalign{\vskip 3pt}
    Datasets & References \\
    \noalign{\vskip 3pt}\hline\noalign{\vskip 1.5pt}\hline\noalign{\vskip 3pt}
    Planck TTTEEE+lowE+lensing & \citep{Planck2018_01} \\
   BICEP/{\it Keck} 2018 (BK18) & \citep{Ade_ea2021} \\
    Baryonic Acoustic Oscillations (BAO) & \citep{Beutler_ea2011, Ross_ea2015, Alam_ea2017} \\
    Dark Energy Survey (DES) & \citep{Abbott_ea2018} \\
    Pantheon+ & \citep{Scolnic_ea2022} \\
    \noalign{\vskip 3pt}\hline\noalign{\vskip 1.5pt}\hline\noalign{\vskip 3pt}
\end{tabular}
\end{table*}

\begin{table}\centering
\caption{\label{tab:datacomb} The datasets combination used in our work.}
\begin{tabular}{p{0.4\columnwidth}}
    \noalign{\vskip 3pt}\hline\noalign{\vskip 1.5pt}\hline\noalign{\vskip 3pt}
    \hfil Datasets \\
    \noalign{\vskip 3pt}\hline\noalign{\vskip 1.5pt}\hline\noalign{\vskip 3pt}
    Planck \\
    Planck + BK18 \\
    Planck + BAO \\
    Planck + DES \\
    Planck + Pantheon+ \\
    Planck + BAO + BK18 \\
    Planck + BAO + DES \\
    \noalign{\vskip 3pt}\hline\noalign{\vskip 1.5pt}\hline\noalign{\vskip 3pt}
\end{tabular}
\end{table}

\section{Data Analysis}
\label{sec:dataanalysis}

We conduct a constraint analysis on the models with a variety of observational data utilizing the Markov Chain Monte Carlo~(MCMC) technique (for a recent review see \cite{Speagle2019}) using {\tt CosmoMC} tool \citep{Lewis_Bridle2002}\footnote{https://cosmologist.info/}. {\tt CosmoMC} is an MCMC program for exploring cosmological parameter space usually work in conjunction with {\tt CAMB}\footnote{https://camb.info/}, which calculates the CMB power spectra based on input cosmological parameters.  We analyse the Markov chains using {\tt GetDist} tool \citep{Lewis2019}\footnote{https://getdist.readthedocs.io/} which gives the marginalized joint probability constraints on parameters of our interest.  The {\tt CosmoMC} and {\tt CAMB} codes are modified to incorporate the models by adding the model parameters.  For the extended Starobinsky $R^3$ model, we add $N_{\rm e}$ and $\delta$ as the model input parameters ($\delta = 0$ for the Starobinsky $R^2$ model).  We apply a uniform prior on $N_{\rm e} \in [50., 80.]$ and $\delta \in [-0.0004, 0.0004]$ for the extended Starobinsky $R^3$ model. For the Weyl model, we incorporate $N_{\rm e}$ and $\gamma_{\rm W}$ as additional parameters.  Similarly, we apply a uniform prior on $N_{\rm e} \in [50., 80.]$ and $\gamma_{\rm W} \in [-1.5\times10^{-3}, 1.5\times10^3]$ for Weyl model. The range of our priors is sufficient to encompass the posteriors, as illustrated in \fref{fig:ns-r} and \fsref{fig:all_data_star_plot1}{fig:all_data_weyl_plot2}.

The power spectrum parameters $n_{\rm s}$, $r$, $n_{\rm run}$, $n_{\rm run,run}$, $n_{\rm t}$ and $n_{\rm t, run}$ are derived from \esref{eq:ns_star}--(\ref{eq:ntrun_star}) for the extended Starobinsky model.  Similar to the extended Starobinsky model, the power spectrum parameters are now derived from \esref{eq:weyl_ns}--(\ref{eq:weyl_ntrun}) for the Weyl model.  For each model (Starobinsky and Weyl), we run an analysis with $k_* = 0.002\ {\rm Mpc}^{-1}$ and $k_* = 0.05\ {\rm Mpc}^{-1}$ (See \eref{eq:power_scalar} and \eref{eq:power_tensor}) respectively.  We also give comments on the choice of $k_*$ in \sref{ssec:effect_kpivot}.

We shall provide a concise overview of the data employed in our analysis.  The summary of the datasets and dataset combinations used in our work are displayed in \tref{tab:dataset} and \tref{tab:datacomb}.

\subsection{Planck TTTEEE+lowE+lensing}

Planck 2018 data release~\citep{Planck2018_01} comprises a combination of CMB temperature, polarization and lensing anisotropies. The data has been compressed using 2-point statistics, especially the angular correlation function, expressed in terms of the multipole moments $C_{\ell}$ as the final output.  There are three types of multipole moments in Planck data: $C_\ell^{\rm TT}$, $C_\ell^{\rm TE}$ and $C_\ell^{\rm EE}$. TT, TE and EE denote temperature auto-correlation, temperature-E-mode polarization cross-correlation and E-mode polarization auto-correlation respectively. In addition, the data also provides an estimate of the power spectrum of the lensing potential and extracted from the data using quadratic estimator \citep{Okamoto_Hu2003}.  The likelihood for temperature and polarization anisotropies measurements use different statistical analysis for large-scale data low-multipole (low-$\ell$ for $2 < \ell < 30$) and small-scale data high-multipole (high-$\ell$ for $\ell \geq 30$).  For low-$\ell$ values, the statistical analysis for the temperature anisotropies is based on the {\tt Commander} likelihood code.  The analysis of low-$\ell$ E-mode polarization likelihood is conducted using the {\tt SimAll} EE likelihood code and is labelled as lowE.  For high values of multipoles moments, the labels TT,TE,EE represent the likelihood analysis for $\ell \geq 30$.  For the lensing likelihood analysis the {\tt SMICA} likelihood is used and is labelled as {\tt lensing} in Planck data.  We adhere to the labelling convention from \cite{Planck2018_01} for likelihoods, where TTTEEE+lowE+lensing refers to combination of temperature and polarization likelihoods at both high-$\ell$ and low-$\ell$, including the lensing likelihood.  In our work, we exclusively utilize the TTTEEE+lowE+lensing Planck data; therefore, when we mention Planck data, we are specifically referring to TTTEEE+lowE+lensing.

\subsection{Bicep/{\it Keck} 2018 (BK18)}

The Bicep/{\it Keck} program involves the BICEP (Background Imaging of Cosmic Extragalactic Polarization) instruments working on {\it Keck Array} telescopes. Its objective is to detect the B-mode polarization of the CMB. The sources of B-mode polarization mainly come from the primordial gravitational waves and astrophysical foreground, in particular from our own galaxy \citep{Ade_ea2021}.  However, both sources emit different B-mode polarization spectra and could be distinguished by multi-frequency measurements.  The primordial gravitational wave is generated from the tensor-mode perturbation during the inflation. The BICEP/{\it Keck} data complements the Planck dataset and the combined BICEP/{\it Keck} and Planck dataset improves the constraint on the tensor-to-scalar ratio $r$.  In this work, we employ BICEP/{\it Keck} 2018 dataset denoted by BK18.

\subsection{Baryonic Acoustic Oscillations (BAO)}

The Baryonic Acoustic Oscillations (BAOs) are an imprint of the acoustic waves mediated by baryon-photon plasma during the time of recombination.  It has an oscillatory feature in the matter power spectrum that defines characteristic length scale or a standard ruler.  The ratio between the transverse distance to the radial distance also gives a characteristic angular scales at each redshift.  We employ the compilation BAO dataset provided by the {\tt CosmoMC} package which comprises of the 6dF survey \citep{Beutler_ea2011} at effective redshift $z_{\rm eff} = 0.106$, the SDSS Main Galaxy Sample \citep{Ross_ea2015} at $z_{\rm eff} = 0.15$ and the SDSS III DR12 data \citep{Alam_ea2017} at $z_{\rm eff} = 0.38, 0.51, 0.61$. The BAO datasets are complimentary to Planck data in terms of temporal coverage.

\subsection{Dark Energy Survey (DES)}

The primary goal of the Dark Energy Survey (DES) is to investigate the properties of dark energy through a comprehensive approach that includes the analysis of both galaxy clustering and weak gravitational lensing, utilizing correlation functions. This extensive study involves the mapping of more than 300 million galaxies and over ten thousand galaxy clusters, covering an area of over 5,000 square degrees \citep{Abbott_ea2018}.  The dataset includes the correlation function of cosmic shear, the angular autocorrelation of luminous red galaxies, and the cross-correlation between the shear of source galaxies and the luminous red galaxy.  We employ the DES dataset provided by the {\tt CosmoMC} package.

\subsection{Pantheon+}

Pantheon+ is a compilation program of all distance measurements of spectroscopically confirmed Type Ia supernovae (SNIa) to date. The data comprises of 1701 light curves and distance modulus of SNIa \citep{Scolnic_ea2022}.  The main goal of the project is to achieve high precision measurements of $H_0$ by calibrating with the low-redshift Cepheid variables data from SH0ES program (Supernovae and H0 for the Equation of State of dark energy) \citep{Riess_ea2022}.  In our work, we modified {\tt CosmoMC} and {\tt CAMB} by employing the distance modulus along with the covariance matrix from Pantheon+ for the likelihood analysis.

\section{Results}
\label{sec:results}

In this section, we provide constraints on our models described in \sref{sec:r2r3models} and \sref{sec:Wmodel}.  For each model, an MCMC analysis described in \sref{sec:dataanalysis} is performed with the data combination in \tref{tab:datacomb}.  The standard $\Lambda$CDM model with the same setting is also included for comparison.  We divide our results into the main parameters which include $\Omega_{\rm b} h^2$, $\Omega_{\rm c} h^2$, $100 \theta_{\rm MC}$, $\tau$, $\ln(10^{10} A_{\rm s})$, $S_8$ and $H_0$.  The power spectrum parameters include $n_{\rm s}$, $r$, $n_{\rm run}$, $n_{\rm run,run}$, $n_{t}$ and $n_{\rm t,run}$.
The results for the main parameters are summarized in \tsref{tab:table1_planck}{tab:table1_bao_des} for $k_* = 0.002, 0.05\ {\rm Mpc}^{-1}$ in \sref{ssec:main_params}, the results for the power spectrum parameters are summarized in \tsref{tab:table2_planck}{tab:table2_bao_des} for $k_* = 0.002, 0.05\ {\rm Mpc}^{-1}$.

The marginalized joint 68\% and 95\% probability regions for the main parameters, $k_* = 0.002\ {\rm Mpc}^{-1}$, are shown in \fref{fig:all_data_star_plot1}, \fref{fig:all_data_exst_plot1} and \fref{fig:all_data_weyl_plot1} in \sref{ssec:main_params} for the Starobinsky $R^2$, $R^3$ and Weyl model respectively.  Similarly, the marginalized joint 68\% and 95\% probability regions for the main parameters, $k_* = 0.05\ {\rm Mpc}^{-1}$, are shown in \fref{fig:all_data_star5_plot1}, \fref{fig:all_data_exst5_plot1} and \fref{fig:all_data_weyl5_plot1} in \sref{ssec:main_params}.  The marginalized joint 68\% and 95\% probability regions for the power spectrum parameters are also shown in \fref{fig:all_data_star_plot2}, \fref{fig:all_data_exst_plot2} and \fref{fig:all_data_weyl_plot2} for $k_* = 0.002\ {\rm Mpc}^{-1}$ in \sref{ssec:ps_params}.

\begin{figure*}
\includegraphics[width=\textwidth]{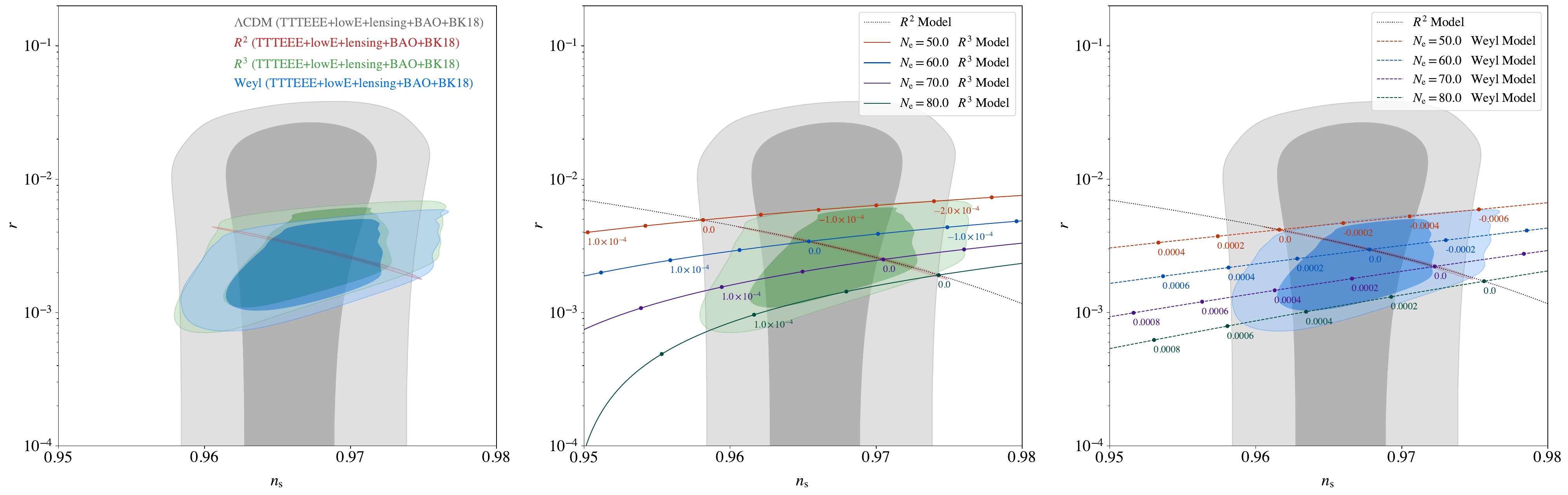}
\caption{\label{fig:ns-r} (left) Marginalized joint 68\% and 95\% CL regions for $n_{\rm s}$ and $r$ ($k_{*} = 0.002$ Mpc$^{-1}$) for $\Lambda$CDM model (grey), Starobinsky $R^2$ model (red), Extended Starobinsky $R^3$ model (green) and Weyl model (blue) from {\it Planck} TTTEEE+lowE+lensing+BAO+BK18. (middle) Same as the left panel but focusing on the Starobinsky and the Extended Starobinsky model with plots of varying $N_{\rm e}$ for $R^2$ model (dotted) and ($N_{\rm e}$, $\delta$) for $R^3$ model (solid) superimposed.  (right) Same as the left panel but focusing on the Starobinsky and the Weyl model with plots of varying $N_{\rm e}$ for $R^2$ model (dotted) and ($N_{\rm e}$, $\gamma_{\rm W}$) for Weyl model (dashed) superimposed.}
\end{figure*}

We also display the constraints on the parameter that are specific to each model ($N_{\rm e}$, $\delta$ or $\gamma_{\rm W}$) in \tref{tab:specific_params} for $k_* = 0.002\ {\rm Mpc}^{-1}$ and $k_* = 0.05\ {\rm Mpc}^{-1}$ respectively.  In additional, we plot the marginalized 68\% and 95\% constraint on $n_{\rm s}$ and $r$ for each model superimposed with standard $\Lambda$CDM constraint from {\it Planck} TTTEEE+lowE+lensing+BAO+BK18 dataset in \fref{fig:ns-r}.  The plots with different datasets resemble the one presented in this article.

\section{Discussions and Conclusions}
\label{sec:discussion}

We shall begin our discussion section by addressing the common features shared by all models before providing a detailed discussion of each individual model.  Additionally, we will offer comparison between different datasets.  We also provide comments on the choice of $k_*$, $H_0$ and $S_8$ tension, constraints on $N_{\rm e}$ from reheating and additional observational data that could further refine the constraints on the model.  Finally, we offer a concluding summary at the end.

\subsection{Common features}

Overall, all the models are in good agreement with all observational data we employed, including both CMB and large-scale structure data, within a constrained range of parameters that are specific to each model.  The common parameter to all the model is the $e$-folding number $N_{\rm e}$.  Only $N_{\rm e}$ in the Starobinsky $R^2$ model could be constrained within a range of approximately $60 - 70$ (See \tref{tab:specific_params} for example) consistent with predictions for other inflationary models (see, for example, \citep{Liddle_ea1994, Groot_vanTent2002, Liddle_Leach2003}).  For the other two models, the values of $N_{\rm e}$ could not be constrained within the range of our prior due to degeneracy with the other parameter ($\delta$ for $R^3$ Starobinsky model or $\gamma_{\rm W}$ for Weyl model).  The Starobinsky $R^2$ model gives tightest constraints overall in comparison to the other models due to having less parameter.  The values of $\delta$ and $\gamma_{\rm W}$ fall within $1\sigma$ from zero.  From the datasets, it is suggested that the faster the inflation ends, the higher the tensor mode. This can be understood from the formulas for $n_{\rm s}$ and $r$ as decreasing functions of $N_{\rm e}$, as given in \eref{eq:ns_star} and \eref{eq:r_star} for the $R^3, R^{2}~(\delta = 0)$ model and \eref{nsWeyl} and \eref{rWeyl} for the Weyl model.

Our constraints on $N_{\rm e}$ for the Starobinsky $R^2$ model differ significantly from the one in \cite{Planck2018_10}. Several reasons account for these discrepancies. The primary factor contributing to this difference is the choice of prior. In \cite{Planck2018_10}, the prior is set to $[50, 60]$, whereas in our work, it is extended to $[50, 80]$.  Another reason is the exact formula we use for $n_{\rm s}$ in \eref{eq:ns_star} which contains higher order terms of $1/N_{\rm e}$, see \eref{eq:ns_Ne} for approximation up to the order $N_{\rm e}^{-2}$.  The inclusion of the higher order terms of $1/N_{\rm e}$ leads to a preference for a higher value of $n_{\rm s}$ from the data, as illustrated in the middle panel of \fref{fig:ns-r}.

For the cosmological parameters, there is good agreement with minor deviations among different models and datasets. The values of the spectral index $n_{\rm s}$ align with those of the standard $\Lambda$CDM. The tensor-to-scalar ratio $r$ can be constrained with both an upper and lower bound. However, the lower bound constraint on $r$ is purportedly influenced by our choice of prior on $N_{\rm e}$ which could go to smaller values of $r$ as $N_{\rm e}$ becomes larger for both the $R^3$ and Weyl models. The additional parameters $\delta$ and $\gamma_W$ compensate for the effect of $N_{\rm e}$, make $r$ smaller while keeping $n_{\rm s}$ within the allowed region from observations. The upper bound constraint on $r$ is also mildly effected by the choice on lower bound on prior of $N_{\rm e}$. We will elaborate this point further in \sref{ssec:r2_r3_discussion} and \sref{ssec:weyl_discussion}.  In addition, we find no strong evidence for the existence of running parameters within the models. However, our results show a preference for a negative value of $n_{\rm run}$ and positive value of $n_{\rm run,run}$ with a deviation within $1\sigma-2\sigma$.  Regarding $n_{\rm t}$ and $n_{\rm t, run}$, our results also suggest a preference for a negative value within $1\sigma-2\sigma$.

\begin{table*}\centering
\caption{\label{tab:specific_params} 95\% CL parameter constraints on parameters specific to the models for $k_* = 0.002, 0.05\ {\rm Mpc}^{-1}$. $N_{\rm e}$ is the parameter common to all models; however, only $N_{\rm e}$ from $R^2$ model could be constrained from the datasets within the range defined by our prior.}
\noindent\begin{tabular}{p{0.22\textwidth}p{0.085\textwidth}p{0.085\textwidth}p{0.085\textwidth}p{0.085\textwidth}p{0.085\textwidth}p{0.085\textwidth}}
	\noalign{\vskip 3pt}\hline\noalign{\vskip 1.5pt}\hline\noalign{\vskip 3pt}
	& \multicolumn{2}{c}{\bf\boldmath$R^2$ Model} & \multicolumn{2}{c}{\bf {\boldmath$R^3$} Model } & \multicolumn{2}{c}{\bf Weyl Model} \\[-0.6em]
	\noalign{\vskip 7pt}\cline{2-7}\noalign{\vskip 3pt}
	{\bf Parameter} 	&  \multicolumn{2}{c}{\boldmath$N_{\rm e}$} & \multicolumn{2}{c}{\boldmath$\delta\ (\times 10^{-4})$} &  \multicolumn{2}{c}{\boldmath$\gamma_{\rm W}\ (\times 10^{-4})$} \\
	\noalign{\vskip 1pt}
	\bf\boldmath$k_*$ (Mpc$^{-1}$) & \hfil\bf 0.002 & \hfil\bf 0.05 & \hfil\bf 0.002 & \hfil\bf 0.05 & \hfil\bf 0.002 & \hfil\bf 0.05 \\
	\noalign{\vskip 3pt}\hline\noalign{\vskip 1.5pt}\hline\noalign{\vskip 5pt}
	{\bf Planck} 		& \hfil $64^{+10}_{-10}$ & \hfil $62^{+10}_{-10}$ & \hfil $-0.1^{+1.3}_{-1.4}$ & \hfil $-0.1^{+1.1}_{-1.3}$ & \hfil $1.1^{+4.5}_{-4.7}$ & \hfil $1.6^{+4.0}_{-4.5}$ \\[0.4em]
	{\bf Planck+BK18} 	& \hfil $63^{+10}_{-10}$ & \hfil $60^{+10}_{-10}$ & \hfil $0.0^{+1.2}_{-1.4}$ & \hfil $0.2^{+1.1}_{-1.3}$ & \hfil $1.1^{+4.4}_{-4.9}$ & \hfil $1.6^{+4.0}_{-4.5}$ \\[0.4em]
	{\bf Planck+BAO} 	& \hfil $67^{+10}_{-10}$ & \hfil $64^{+10}_{-10}$ & \hfil $-0.2^{+1.2}_{-1.4}$ & \hfil $-0.1^{+1.0}_{-1.2}$ & \hfil $0.4^{+4.2}_{-4.8}$ & \hfil $1.0^{+3.9}_{-4.2}$ \\[0.4em]
	{\bf Planck+DES}	& \hfil $70^{+10}_{-10}$ & \hfil $67^{+10}_{-10}$ & \hfil $-0.4^{+1.2}_{-1.4}$ & \hfil $-0.3^{+1.1}_{-1.3}$ & \hfil  $-0.1^{+4.3}_{-4.8}$ & \hfil $0.5^{+4.0}_{-4.4}$ \\[0.4em]
	{\bf Planck+Pantheon+} & \hfil $65^{+10}_{-10}$ & \hfil $63^{+10}_{-10}$ & \hfil $-0.2^{+1.2}_{-1.4}$ & \hfil $0.0^{+1.1}_{-1.3}$ & \hfil $0.7^{+4.3}_{-5.0}$ & \hfil $1.3^{+3.8}_{-4.4}$ \\[0.4em]
    {\bf Planck+BAO+BK18}  & \hfil $66^{+10}_{-10}$ & \hfil $63^{+10}_{-10}$ & \hfil $-0.2^{+1.2}_{-1.4}$ & \hfil $-0.2^{+1.1}_{-1.2}$ & \hfil $0.2^{+4.3}_{-4.7}$ & \hfil $0.9^{+3.7}_{-4.2}$ \\[0.4em]
	{\bf Planck+BAO+DES} & \hfil $70^{+10}_{-10}$ & \hfil $67^{+10}_{-10}$ & \hfil \hfil $-0.4^{+1.2}_{-1.5}$ & \hfil $0.0^{+1.1}_{-1.2}$ & \hfil  $-0.1^{+3.9}_{-4.9}$ & \hfil $0.4^{+3.6}_{-4.3}$ \\
	\noalign{\vskip 3pt}\hline\noalign{\vskip 1.5pt}\hline\noalign{\vskip 5pt}
\end{tabular}
\end{table*}

\subsection{$R^2$ Model and $R^3$ Model}
\label{ssec:r2_r3_discussion}

The primary distinction between the Starobinsky $R^2$ model and the extended Starobinsky $R^3$ model lies in the value of the $\delta$ parameter in \esref{eq:ns_star}--(\ref{eq:ntrun_star}).  The Starobinsky $R^2$ model is obtained by setting $\delta = 0$ within the extended $R^3$ model.  For the $R^3$ model, the value of $\delta$ is close to zero and falls within the range of $-2 \times 10^{-4} < \delta < 1 \times 10^{-4}$. However, the probability density function (pdf) is negatively skewed to the left, indicating a preference for negative values.  The negative values of $\delta$ would suggest a preference for higher values of $n_{\rm s}$ and $r$ as shown in \fref{fig:ns-r} as well as lower value of $N_{\rm e}$.

\fref{fig:all_data_star_plot1} displays the marginalized posterior probability distribution for the main parameters, while \fref{fig:all_data_star_plot2} presents the same plot for the power spectrum parameters of the Starobinsky $R^2$ model. Similarly \fref{fig:all_data_exst_plot1} and \fref{fig:all_data_exst_plot2} displays the same plot for the extended Starobinsky $R^3$ model.  Our results are in good agreement with \cite{Cheong_ea2020}.  The power spectrum parameters $n_{\rm s}$, $r$, $n_{\rm run}$, $n_{\rm run,run}$, $n_{\rm t}$ and $n_{\rm t, run}$ in $R^2$ Starobinsky model are strongly correlated to one another due to explicit relations in \esref{eq:ns_star}--(\ref{eq:ntrun_star}). Nevertheless, the correlations of the power spectrum in the $R^3$ Starobinsky model are less pronounced, mainly because of the influence of the $\delta$ parameter.  

For $R^3$ model, $N_{\rm e}$ is less constrained due to the presence of the additional parameter $\delta$ as mentioned in the aforementioned section. From (\eref{eq:r_Ne}), for large $N_{\rm e}$, $r$ could go arbitrarily small while $n_{\rm s}$ could still lie within the constraints from observations. This renders the model dependent on the prior of $N_{\rm e}$ and $\delta$.  Hence, an independent prior on $N_{\rm e}$ is crucial for constraining the $R^3$ model.

\subsection{Weyl Model}
\label{ssec:weyl_discussion}

The Weyl model differs from the Starobinsky model in its origin by incorporating an additional scalar field instead of additional terms in the geometrical part in the gravity action.  However, the Weyl model exhibits many features that are similar to the Starobinsky model, especially the $R^3$ model.  For example, the dependence on an additional parameter apart from $N_{\rm e}$, leading to constraints on the power spectrum that are less pronounced than those of the $R^2$ model.  The constraining power on the parameters from the Weyl model is also similar to that of the $R^3$ model (also $R^2$ model for the main cosmological parameters), as explicitly seen in \fref{fig:ns-r}.  Due to the presence of additional parameter $\gamma_{\rm W}$, the Weyl model also has less constrained nature in $N_{\rm e}$--though not as explicitly as the $R^3$ model--due to its dependence on $N_{\rm e}$ through the function $\Theta$. As $N_{\rm e}$ grows, the Weyl model tends to prefer smaller values of $r$ and a higher value of $n_{\rm s}$, similar to the $R^3$ model. For large $N_{\rm e}$, turning on small values of $\gamma_{\rm W}$ could keep $n_{\rm s}$ within the allowed region from observations. The allowed value of $\gamma_{\rm W}$ is very close to zero and lies within the range $-5 \times 10^{-4} < \gamma_{\rm W} < 5 \times 10^{-4}$ as shown in \tref{tab:specific_params}. The posterior probability density function of $\gamma_{\rm W}$ is slightly positively skewed, indicating preference on positive values as well as higher value of $N_{\rm e}$ and lower value of $n_{\rm s}$ and $r$. 

\subsection{Comparison between datasets}

We conducted assessments of our models, utilizing data from CMB sources (Planck, BK18) and large-scale structure datasets (BAO, Pantheon+, DES), covering various cosmological parameters. Planck served as the foundational dataset, supplemented with additional data (See \tref{tab:datacomb}).  Therefore, the main constraining power typically arises from the Planck CMB data.  In general, the constraints on cosmological parameters are similar to one another but exhibit some consistent deviations between datasets. For example, the CMB data (Planck and BK18) consistently favours lower values of $\Omega_{\rm b}h^2$ than the large-scale data, whereas the opposite is true for $\Omega_{\rm c}h^2$ as can be seen from \tsref{tab:table1_planck}{tab:table1_bao_des}. From the tables, the values of $\theta_{\rm MC}$ and $\tau$ also consistently increase across different datasets, from CMB to large-scale structure data while the opposite is true for $A_{\rm s}$.  Regarding the power spectrum parameters, there is a tendency for the values of $n_{\rm s}$ to increase from CMB data to large-scale data as can be observed from \tsref{tab:table2_planck}{tab:table2_bao_des}.  In general, the variation in the constraints on cosmological parameters consistently differs between the CMB data (Planck, BK18) and large-scale structure data (BAO, Pantheon+, DES). The extreme constraints from the CMB data are from BAO+BK18 or BK18, while the other extreme arises from DES or BAO+DES.

\subsection{Effects of $k_*$ on Cosmological Parameters}
\label{ssec:effect_kpivot}

In this work, the two choices of $k_{*}$; $k_{*} = 0.002\ {\rm Mpc}^{-1}$ and $k_{*} = 0.05\ {\rm Mpc}^{-1}$, are used as benchmarks in our work for the purpose of comparison with existing literature (for example, \cite{Planck2018_10,Planck2018_06}). The choice of $k_{*}$ is arbitrary; however, it is often based on practical considerations and the goal of capturing relevant information specific to a model or dataset.  Our results show no notable distinction in the constraints on cosmological parameters between the two choices of $k_{*}$ except for $N_{\rm e}$, $\delta$, $\gamma_{\rm W}$ and $A_{\rm s}$.  It is worth noting, as shown in \tref{tab:specific_params}, that the mean value of $N_{\rm e}$ differs by 2--3 between the two choices of $k_*$.  Typically, $k_* = 0.002\ {\rm Mpc}^{-1}$ favours lower values of $\delta$ and $\gamma_{\rm W}$ in comparison to $k_* = 0.05\ {\rm Mpc}^{-1}$.

Referring to \eref{eq:power_scalar}, for the scalar power spectrum amplitude $A_{\rm s}$, the values of $\ln A_{\rm s}$ at two different $k_*$ values are well approximated by
\begin{eqnarray}
    \nonumber \ln\Big(A_{\rm s}\;\big\vert_{k_* = \rm 0.002\ Mpc^{-1}} \Big) &\approx& \ln\Big(A_{\rm s}\;\big\vert_{k_* = \rm 0.05\ Mpc^{-1}} \Big) \\
    && + \ln 25 \times (1 - n_{\rm s}),
\end{eqnarray}
where $A_{\rm s}\;\big\vert_{k_* = \rm 0.002\ Mpc^{-1}}$ and $A_{\rm s}\;\big\vert_{k_* = \rm 0.05\ Mpc^{-1}}$ are the scalar amplitude at $k_* = 0.002\ {\rm Mpc}^{-1}$ and $k_* = 0.05\ {\rm Mpc}^{-1}$ respectively.

\begin{figure}\centering
\includegraphics[width=0.5\columnwidth]{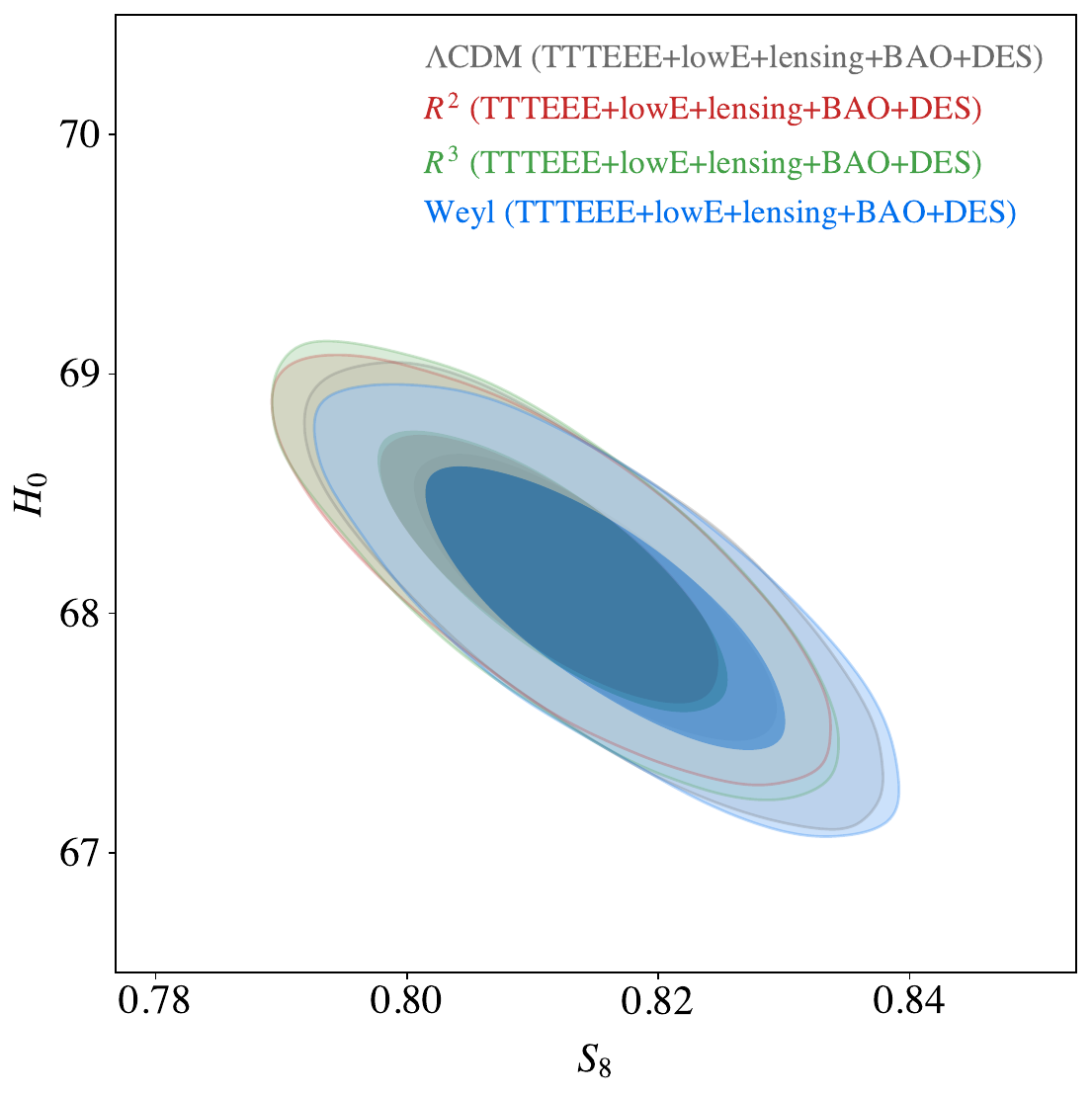}
\caption{\label{fig:sigma8-h0} Marginalized joint 68\% and 95\% CL regions for $H_0$ and $S_8$ ($k_{*} = 0.05$ Mpc$^{-1}$) for Planck+BAO+DES between different models.}
\end{figure}

\subsection{$H_0$ and $S_8$ Tensions}
\label{ssec:h0_sigma8_tension}

The Hubble tension stands out as one of the most statistically significant discrepancies in observational cosmology, showing a disagreement of 4$\sigma$ to 6$\sigma$ between early-time and late-time observations (See~\cite{DiValentino_ea2021} and references therein). For example, a late-time measurement of $H_0$ from Pantheon+SH0ES yields $H_0 = 73.04 \pm 1.04\ {\rm km\  s}^{-1}\ {\rm Mpc}^{-1}$ \citep{Riess_ea2022}, while an early-time measurement from Planck CMB gives $H_0 = 67.4 \pm 0.5\ {\rm km\ s}^{-1}\ {\rm Mpc}^{-1}$ \citep{Planck2018_06}. The discrepancy between these two measurements is approximately $5.5\sigma$.

Another tension related to the inconsistency between early-time and late-time measurements involves $\sigma_{8}$, which represents the amplitude of the power spectrum at a scale of 8 Mpc. This tension is commonly expressed in terms of $S_8$ ($\sigma_8 (\Omega_{\rm m}/0.3)^{0.5}$), influencing the amplitude of weak lensing measurements. The measurements from lower redshift probes systematically favour lower value of $S_8$ compared to those obtained from high-redshift CMB measurements \citep{Abdalla_ea2022}. For instance, an early-time measurement from Planck CMB gives $S_8 = 0.834 \pm 0.016$ \citep{Planck2018_06}, while a late-time measurement from DES weak gravitational lensing yields $S_8 = 0.759 \pm 0.025$ \citep{Amon_ea2022}.

When comparing constraints across different datasets, we observe that the Planck+DES and Planck+BAO+DES datasets systematically favour a lower value of $S_8$ and a higher value of $H_0$.  For example, the constraints on $(S_8, H_0)$ for $R^2$ model are $(0.812^{+0.018}_{-0.018}, 68.19^{+0.72}_{-0.72})$ for Planck+BAO+DES, while Planck alone gives $(0.830^{+0.024}_{-0.024}, 67.50^{+1.00}_{-0.99})$.  When comparing between different models, our findings indicate that, with the exception of Planck + DES and Planck + BAO + DES dataset, the constraints on $S_8$ and $H_0$ remain consistent across all datasets and all models. \fref{fig:sigma8-h0} displays marginalized joint 68\% and 95\% confidence level regions for $S_{8}$ and $H_0$ ($k_{*} = 0.05$ Mpc$^{-1}$) for Planck+DES amongst different models. In terms of mean values, the Starobinsky models exhibit a tendency toward lower values of $S_8$ and higher values of $H_0$ compared to the $\Lambda$CDM and Weyl model. Our results continue to highlight tension between Planck CMB measurements and late-time observations, particularly with respect to the DES dataset. However in \cite{Nunes_Vagnozzi2021}, it is demonstrated that when viewed through the lens of growth rate measurements from Redshift Space Distortion (RSD) datasets, the $S_8$ discrepancy could be compatible with a statistical fluctuation. Consequently, its significance might be overestimated.

\subsection{Reheating Constraints on $N_{\rm e}$}
\label{ssec:reheating_con}

 The $e$-folding number required to solve the horizon problem, i.e., the $e$-folding number from the horizon-exit to the end of inflation $N_{\rm e}$, depends on the particle content of the universe and the temperature during the reheating by the relation~\cite{Liddle_Leach2003,Martin:2010kz,Planck:2015sxf}
\begin{eqnarray}
N_{\rm e}&\simeq& 67-\ln \left(\frac{k_{*}}{a_{0}H_{0}}\right)+\frac{1}{4}\ln \left(\frac{V^{2}_{*}}{M^{4}_{\rm P}\rho_{\rm end}}\right) + \frac{1-3w_{\rm int}}{12(1+w_{\rm int})}\ln\left( \frac{\rho_{\rm th}}{\rho_{\rm end}}\right)
- \frac{1}{12}\ln(g_{\rm th}),    \label{Np:eq}
\end{eqnarray}
where $(a_{0}H_{0})^{-1}$ is the present comoving Hubble length and $k_{*}$ is the horizon-exit scale.

\begin{figure}[h!]
\centering
\includegraphics[width=0.7\textwidth]{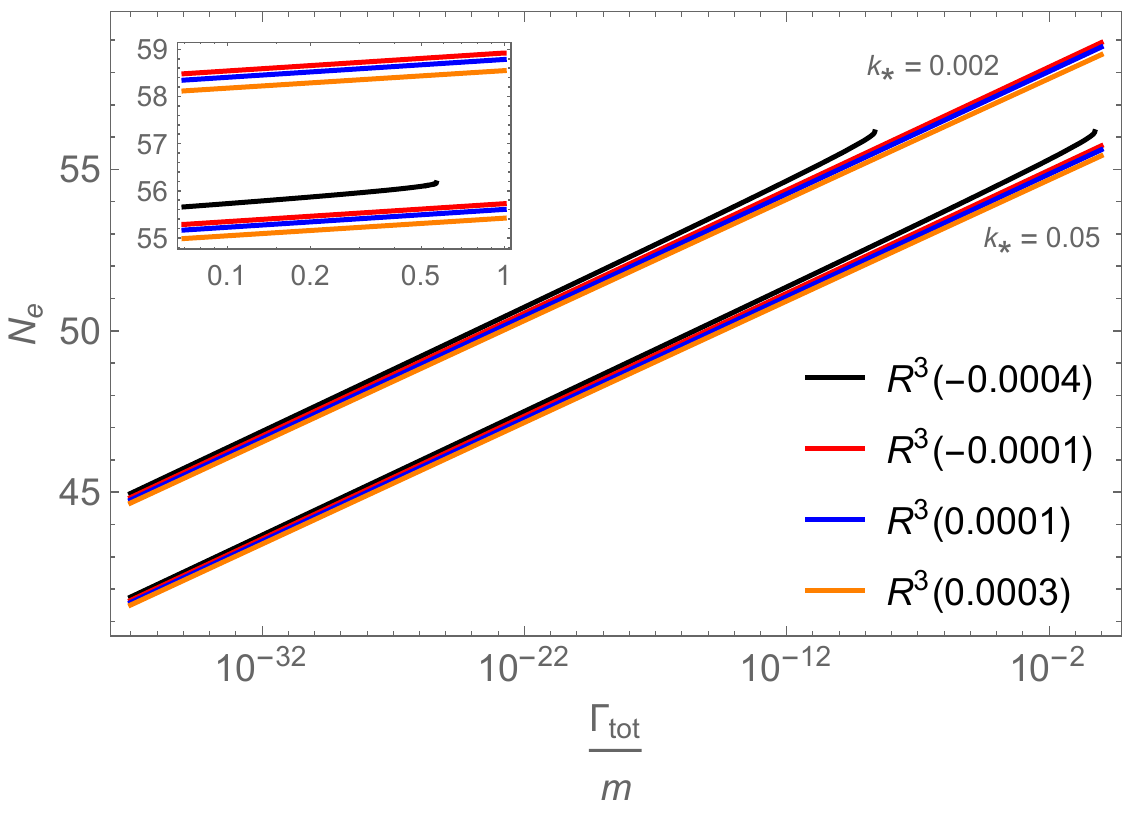}
\caption{\label{fig:NR3reh} Plot between the required $e$-folding number from horizon exit to the end of inflation $N_{\rm e}$ versus $\Gamma_{\rm tot}/m$ in extended Starobinsky model for $\delta=-0.0004,-0.0001,0.0001,0.0003$ and $k_{*}=0.002, 0.05$ Mpc$^{-1}$, $g_{\rm th}=106.75$ is assumed for the Standard Model particle production.}
\end{figure}

To account for the effect of reheating decay of inflaton to the standard model particles in each model, we assume the total decay rate of inflaton to be $\Gamma_{\rm tot}$. In the extended Starobinsky model, the inflaton mass can be calculated from $V(s)\simeq V_{0}(s)$
\begin{equation}
V_{0}(s)\simeq \frac{s^{2}}{6\beta}M^{2}_{\rm P},
\end{equation}  
therefore the scalar mass is
\begin{equation}
m_{s}=\frac{M_{\rm P}}{\sqrt{3\beta}},   
\end{equation}
where we retrieved the reduced Planck mass $M_{\rm P}$. Assuming reheating with $w_{\rm int}=0$ and the approximation $V_{\rm end}\approx V_{*}$, \eref{Np:eq} can be expressed as the following
\begin{eqnarray}
N_{\rm e}&\simeq& 67-\ln \left(\frac{k_{*}}{a_{0}H_{0}}\right)+\frac{1}{6}\ln\left( \frac{\Gamma_{\rm tot}}{m}\frac{V_{*}}{\sqrt{3\beta}M_{\rm P}^{4}}\right)  \notag \\
&-& \frac{1}{12}\ln(g_{\rm th}/3).    \label{NpR3:eq}
\end{eqnarray}
We have used the relation $\rho_{\rm th}=g_{\rm th}\displaystyle{\frac{\pi^{2}T_{\rm reh}^{4}}{30}}$ for reheating temperature saturating the upper bound~\cite{Bassett:2005xm}
\begin{equation}
T_{\rm reh}=\left( \frac{90}{g_{\rm th}\pi^{2}}\right)^{1/4}\sqrt{\Gamma_{\rm tot}M_{\rm P}}.    
\end{equation}
By using the relation of $\beta$ and $N_{\rm e}$ from \eref{be:eq}, we can numerically plot $N_{\rm e}$ versus $\Gamma_{\rm tot}/m$ for the extended Starobinsky model for $k_{*}=0.002, 0.05$ Mpc$^{-1}$ as presented in \fref{fig:NR3reh}.

\begin{figure}[h!]
\centering
\includegraphics[width=0.7\textwidth]{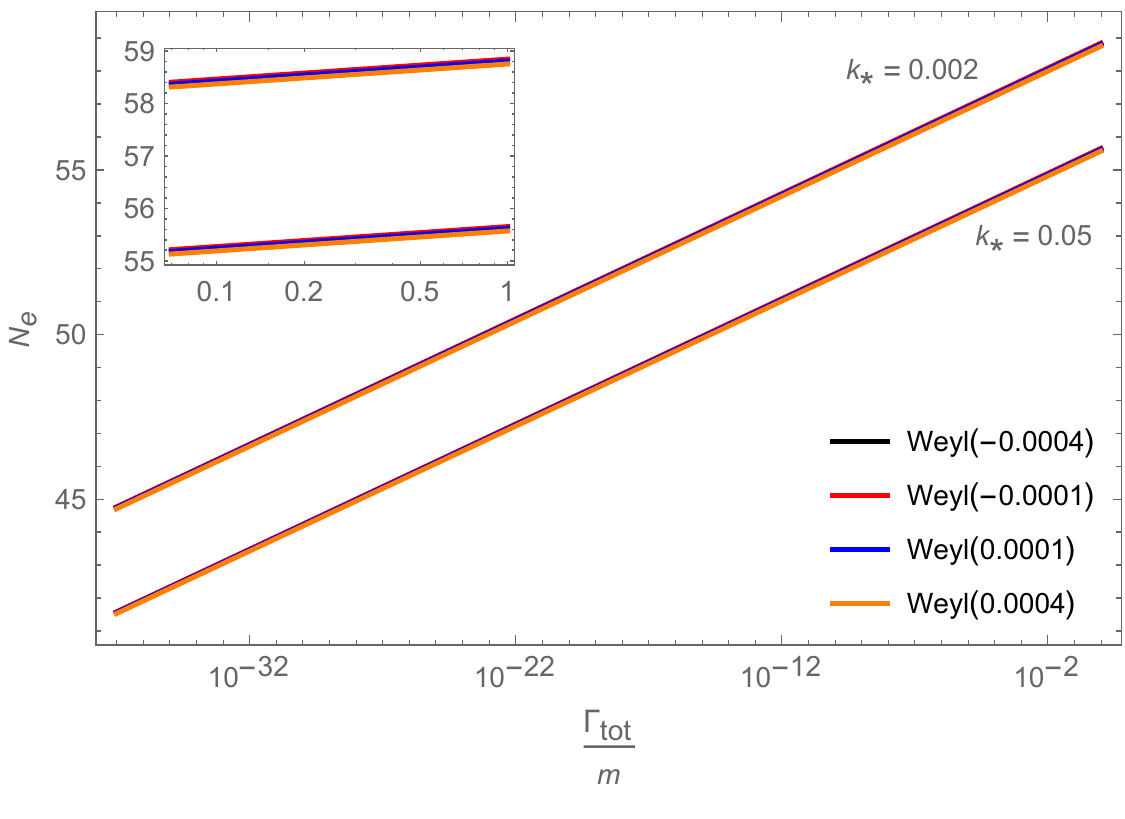}
\caption{\label{fig:Weylreh} Plot between the required $e$-folding number from horizon exit to the end of inflation $N_{\rm e}$ versus $\Gamma_{\rm tot}/m$ in Weyl gravity model for $\gamma_{\rm W}=-0.0004,-0.0001,0.0001,0.0004$ and $k_{*}=0.002, 0.05$ Mpc$^{-1}$, $g_{\rm th}=106.75$ is assumed for the Standard Model particle production.}
\end{figure}

For Weyl inflationary model, the potential (\eref{VWeq}) gives the mass term
\begin{equation}
V_{\rm E}(\Phi)\simeq \frac{(\Phi - \Phi_{0})^{2}M^{2}_{\rm P}}{12\alpha},
\end{equation}
which implies $m_{\Phi}=\displaystyle{\frac{M_{\rm P}}{\sqrt{6\alpha}}}$. This leads to the same result as the extended Starobinsky model with a replacement $\beta \to 2\alpha$ in \eref{NpR3:eq} and
\begin{equation}
\frac{V_{*}}{\epsilon(\Phi_{*})}=24\pi^{2}A_{\rm s}M^{4}_{\rm P}\simeq 0.027^{4}M^{4}_{\rm P}, \label{WVs:eq}
\end{equation}
where we used $\ln(10^{10}A_{\rm s})\simeq 3.10$ from \cite{Planck2018_10}. Similar to the extended Starobinsky model, the relation \eref{WVs:eq} gives the value of $\alpha$ as a function of $N_{\rm e}$. At each value of $\Gamma_{\rm tot}/m$, $N_{\rm e}$ can be numerically determined as shown in \fref{fig:Weylreh}.

A few remarks are in order. For extended Starobinsky model, there is a truncation of plot at around $N_{\rm e}\simeq 56$ for $\delta = -0.0004$, i.e., there is no solution satisfying \eref{NpR3:eq} when the decay rate per mass $\Gamma_{\rm tot}/m$ is too large. On the other hand for $\delta = 0.0004, k_{*}=0.002$ Mpc$^{-1}$, the solution for large $\Gamma_{\rm tot}/m$ converges to $N_{\rm e}\simeq 57$, this is however not shown in \fref{fig:NR3reh} where we choose to present the curve for $\delta = 0.003$ instead. For Weyl model, the dependence of $N_{\rm e}$ on the model parameter $\gamma_{\rm W}$ is less distinctive. Remarkably, reheating constraints set the range of upper-bound saturating $N_{\rm e}$~($\Gamma_{\rm tot}/m \lesssim 1$) to $N_{\rm e} = 55-59$ for both models. Even when all observational data allows very high values of $N_{\rm e}>60$ as depicted in \fref{fig:ns-r}, reheating decay of inflaton to Standard Model particles could set the upper bound on $N_{\rm e}<59$ for both extended Starobinsky and Weyl models.  We note that our conclusion generally follows the constraint from Planck CMB data; however, the constraint on $N_{\mathrm{e}}$ may vary with different datasets, such as ACTPol \cite{Giare_ea2023}.

\subsection{Future Observational Constraints}
\label{ssec:future_obs}

In the forthcoming years, we would be able to evaluate inflationary predictions through direct measurements of the tensor power spectrum. This will primarily involve detecting B-mode polarization arising from gravitational waves generated during inflation, commonly known as primordial gravitational waves. The significance of detecting primordial gravitational waves cannot be overemphasized, as they hold important information about the physics of the very early Universe. Confirming the inflationary scenarios becomes pivotal, as the detection of the tensor-to-scalar ratio $r$ can be directly inferred the energy scale of inflation \citep{Lyth_Riotta1999}.  This assessment will encompass both current and upcoming experiments. For a comprehensive overview of experiments targeting the measurement of primordial gravitational waves, we suggest \cite{Campeti_ea2021}. 

According to \fref{fig:ns-r}, the absence of detection of $r$ above 0.001 would possibly lead to exclusion of the Starobinsky $R^2$ model and impose stringent constraints on the Starobinsky $R^3$ model and Weyl model; hence, we shall discuss experiments that could potentially provide stringent constraints or a definitive rejection of the models in the next decade. For example, the Lite (Light) satellite for the study of B-mode polarization and inflation from cosmic background Radiation Detection (LiteBIRD)~\citep{LiteBIRD-Collaboration2023} is a space-based experiment which will study the B-mode polarization from CMB.  It aims to establish a lower limit on the tensor-to-scalar ratio.  The expected sensitivity for LiteBIRD on tensor-to-scalar ratio is $r < 0.003$ at 95\% confidence level for a fiducial model with $r = 0$.  Tighter constraints are anticipated when combining LiteBIRD data with that from other experiments~\citep{Crowder_Cornish2005}. Similarly, Simons Observatory (SO), which is a ground-based CMB experiment, is also anticipated to give $r < 0.003$ at 95\% confidence level for a fiducial model with $r = 0$ \citep{Ade_ea2019}.

Looking far into the coming decades, planned space-based experiments include the DECi-hertz Interferometer Gravitational wave Observatory (DECIGO)~\citep{Seto_ea2001, Kuroyanagi_ea2015}, Big Bang Observer (BBO)~\citep{Crowder_Cornish2005}, $\mu$ARES~\citep{Sesana_ea2021} which are laser interferometers similar to Laser Interferometer Space Antenna (LISA) \citep{Bartolo_ea2016}.  Those proposed experiments aim to push the constraint on $r$ to the cosmic variance limit, achieving $r < 10^{-4}$ for a fiducial model with $r = 0$.

Here is the summary of our results.
\begin{itemize}
    \item We consider three different inflationary models from different theories of gravitation; Starobinsky $R^2$, extended Starobinsky $R^3$ and Weyl model by considering scalar perturbations and tensor perturbations parameterized by power-law forms in \eref{eq:power_scalar} and \eref{eq:power_tensor}.  The constraints on cosmological parameters, derived using the dataset in \tref{tab:dataset}, indicate that all models are in good agreement with the observational data.
    \item Only $N_{\rm e}$ in the Starobinsky $R^2$ model could be constrained giving the mean value of $N_{\rm e}$ approximately 60-70 consistent with predictions from other inflation models (See \tref{tab:specific_params}).
    \item While the Weyl model differs from the extended Starobinsky $R^3$ model in its origin, the observational constraints on both models are very similar. Hence, distinguishing between the models would require independent observations. However, with current datasets that cannot probe very small tensor-to-scalar ratio $r<0.002$ region, there is no compelling evidence supporting the preference of the $R^3$ model and the Weyl model over the $R^{2}$ Starobinsky model. Future observations would allow us to distinguish the $R^{2}$ model from the $R^3$ and Weyl model with the probe in $r<0.002$ region of the parameter space.
    \item We investigate the effect of the choice of $k_*$ that are frequently used in literature and find that $k_* = 0.002\ {\rm Mpc}^{-1}$ favours lower value of $\delta$ and $\gamma_{\rm W}$ compared to $k_* = 0.05\ {\rm Mpc}^{-1}$. For all models, larger $k_*$ results in smaller constrained value of $A_{\rm s}$. In addition, the mean value of $N_{\rm e}$ differs by 2--3 between the two choices of $k_*$'s.
    \item Our results continue to emphasize  the tension in $H_0$ and $S_8$ between early-time CMB measurements and late-time large-scale structure observations.
    \item Our results indicate that incorporating higher-order terms relaxes the limitation on the upper boundary of the e-folding number $N_{\rm e}$ because of the introduction of small additional parameters. However, the upper limit on $N_{\rm e}$ could be further refined by accounting for the reheating mechanism to $N_{\rm e}<55-59$ for $k_{*}=0.002-0.05$ Mpc$^{-1}$.
\end{itemize}

\section*{Acknowledgments}
We would like to thank Utane Sawangwit and the National Astronomical Research Institute of Thailand~(NARIT) for facilitating the Chalawan HPC and thank greatly the NSTDA Supercomputer center (ThaiSC) and the National e-Science Infrastructure Consortium for their support of computing facilities used in this work. PB and PB are supported in part by National Research Council of Thailand~(NRCT) and Chulalongkorn University under Grant N42A660500. This research has received funding support from the NSRF via the Program Management Unit for Human Resources \& Institutional Development, Research and Innovation~[grant number B39G660025]. TC is supported by Naresuan University (NU), and the National Science, Research and Innovation Fund (NSRF) Grant number R2566B091.

\section*{Data availability}
The data underlying this article will be shared on reasonable request to the corresponding author.

\bibliographystyle{cas-model2-names}
\bibliography{inflation_ref}

\appendix

\section{Appendixes}

In this section, we shall provide all the tables and figures displaying the constraints on relevant parameters.  For detailed explanations and discussions, the reader should refer to the main text in \sref{sec:results} and \sref{sec:discussion}.

\subsection{Parameter Constraints}
\label{sec:param_constr}

\subsubsection{Main parameters}
\label{ssec:main_params}

We provide constraints on the main cosmological parameter that are relevant to our work. The standard parameters are $\Omega_{\rm b} h^2$, $\Omega_{\rm c}h^2$, $\theta_{\rm MC}$, $\tau$, ${\rm ln}(10^{10}A_{\rm s})$.  We also include $H_0$ and $S_8$ in the parameter set as a reference to the discussion on $H_0$ and $S_8$ tension in \sref{ssec:h0_sigma8_tension}.

\begin{table*}[h!]\centering
\begin{tabular}{lcccc}
\noalign{\vskip 3pt}\hline\noalign{\vskip 1.5pt}\hline\noalign{\vskip 5pt}
 &  \multicolumn{1}{c}{\bf\boldmath $R^2$ Model} &  \multicolumn{1}{c}{\bf\boldmath $R^3$ Model} &  \multicolumn{1}{c}{\bf Weyl Model} &  \multicolumn{1}{c}{\bf\boldmath $\Lambda$CDM}\\
\noalign{\vskip 3pt}\cline{2-5}\noalign{\vskip 3pt}

 Parameter &  95\% limits &  95\% limits &  95\% limits &  95\% limits\\[0.4em]
\noalign{\vskip 3pt}\hline\noalign{\vskip 1.5pt}\hline\noalign{\vskip 3pt}

$\Omega_{\rm b} h^2$ & $0.02239^{+0.00028}_{-0.00028}$ & $0.02238^{+0.00028}_{-0.00028}$ & $0.02237^{+0.00029}_{-0.00029}$ & $0.02238^{+0.00030}_{-0.00029}$ \\[0.4em]

& $0.02239^{+0.00028}_{-0.00028}$ & $0.02238^{+0.00029}_{-0.00028}$ & $0.02237^{+0.00030}_{-0.00028}$ & $0.02238^{+0.00029}_{-0.00029}$ \\[0.4em]

$\Omega_{\rm c} h^2$ & $0.1199^{+0.0023}_{-0.0022}$ & $0.1200^{+0.0023}_{-0.0023}$ & $0.1200^{+0.0024}_{-0.0024}$ & $0.1199^{+0.0024}_{-0.0023}$ \\[0.4em]

& $0.1198^{+0.0022}_{-0.0022}$ & $0.1200^{+0.0023}_{-0.0024}$ & $0.1201^{+0.0023}_{-0.0023}$ & $0.1199^{+0.0024}_{-0.0024}$ \\[0.4em]

$100\theta_{\rm MC}$ & $1.04092^{+0.00059}_{-0.00059}$ & $1.04092^{+0.00061}_{-0.00060}$ & $1.04091^{+0.00060}_{-0.00061}$ & $1.04092^{+0.00062}_{-0.00061}$ \\[0.4em]

& $1.04094^{+0.00059}_{-0.00060}$ & $1.04092^{+0.00060}_{-0.00061}$ & $1.04090^{+0.00060}_{-0.00059}$ & $1.04091^{+0.00060}_{-0.00062}$\\[0.4em]

$\tau$ & $0.055^{+0.015}_{-0.014}$ & $0.054^{+0.015}_{-0.015}$ & $0.054^{+0.016}_{-0.015}$ & $0.055^{+0.015}_{-0.014}$ \\[0.4em]

& $0.055^{+0.015}_{-0.014}$ & $0.055^{+0.015}_{-0.014}$ & $0.055^{+0.015}_{-0.014}$ & $0.055^{+0.015}_{-0.015}$ \\[0.4em]

${\rm{ln}}(10^{10} A_{\rm s})$ & $3.155^{+0.032}_{-0.032}$ & $3.155^{+0.034}_{-0.033}$ & $3.156^{+0.033}_{-0.034}$ & $3.155^{+0.033}_{-0.033}$ \\[0.4em]

& $3.046^{+0.029}_{-0.028}$ & $3.045^{+0.029}_{-0.028}$ & $3.045^{+0.029}_{-0.028}$ & $3.045^{+0.030}_{-0.028}$ \\[0.4em]

$H_0$ & $67.40^{+0.99}_{-1.0}$ & $67.4^{+1.1}_{-1.0}$ & $67.3^{+1.1}_{-1.1}$ & $67.4^{+1.1}_{-1.1}$ \\[0.4em]

& $67.5^{+1.0}_{-0.99}$ & $67.4^{+1.1}_{-1.0}$ & $67.3^{+1.1}_{-1.0}$ & $67.4^{+1.1}_{-1.1}$ \\[0.4em]

$S_8$ & $0.831^{+0.024}_{-0.024}$ & $0.831^{+0.025}_{-0.025}$ & $0.832^{+0.026}_{-0.025}$ & $0.831^{+0.026}_{-0.026}$ \\[0.4em]

& $0.830^{+0.024}_{-0.024}$ & $0.831^{+0.025}_{-0.025}$ & $0.832^{+0.025}_{-0.025}$ & $0.831^{+0.025}_{-0.025}$ \\[0.4em]

\noalign{\vskip 3pt}\hline\noalign{\vskip 1.5pt}\hline\noalign{\vskip 3pt}
\end{tabular}
\caption{\label{tab:table1_planck} 95\% Confidence limits of the main cosmological parameters for Planck TTTEEE+lowE+lensing dataset.  The upper set of figures for each parameter are constraints at $k_* = 0.002\ {\rm Mpc}^{-1}$ and the lower set of figures are constraints at $k_* = 0.05\ {\rm Mpc}^{-1}$.}
\end{table*}

\begin{table*}[h!]\centering
\begin{tabular} { l  c c c c}
\noalign{\vskip 3pt}\hline\noalign{\vskip 1.5pt}\hline\noalign{\vskip 5pt}
 &  \multicolumn{1}{c}{\bf\boldmath $R^2$ Model} &  \multicolumn{1}{c}{\bf\boldmath $R^3$ Model} &  \multicolumn{1}{c}{\bf Weyl Model} &  \multicolumn{1}{c}{\bf\boldmath $\Lambda$CDM}\\
\noalign{\vskip 3pt}\cline{2-5}\noalign{\vskip 3pt}

 Parameter &  95\% limits &  95\% limits &  95\% limits &  95\% limits\\[0.4em]
\noalign{\vskip 3pt}\hline\noalign{\vskip 1.5pt}\hline\noalign{\vskip 3pt}
$\Omega_{\rm b} h^2$ & $0.02241^{+0.00028}_{-0.00027}$ & $0.02241^{+0.00029}_{-0.00028}$ & $0.02240^{+0.00029}_{-0.00030}$ & $0.02241^{+0.00028}_{-0.00028}$ \\[0.4em]

& $0.02242^{+0.00029}_{-0.00027}$ & $0.02241^{+0.00029}_{-0.00028}$ & $0.02240^{+0.00028}_{-0.00028}$ & $0.02241^{+0.00029}_{-0.00027}$\\[0.4em]

$\Omega_{\rm c} h^2$ & $0.1196^{+0.0023}_{-0.0022}$ & $0.1196^{+0.0024}_{-0.0024}$ & $0.1197^{+0.0023}_{-0.0023}$ & $0.1195^{+0.0024}_{-0.0024}$ \\[0.4em]

& $0.1195^{+0.0022}_{-0.0022}$ & $0.1196^{+0.0024}_{-0.0023}$ & $0.1197^{+0.0023}_{-0.0023}$ & $0.1195^{+0.0024}_{-0.0024}$\\[0.4em]

$100\theta_{\rm MC}$ & $1.04097^{+0.00060}_{-0.00060}$ & $1.04095^{+0.00060}_{-0.00060}$ & $1.04096^{+0.00059}_{-0.00063}$ & $1.04097^{+0.00062}_{-0.00062}$ \\[0.4em]

& $1.04098^{+0.00059}_{-0.00059}$ & $1.04095^{+0.00060}_{-0.00060}$ & $1.04095^{+0.00059}_{-0.00060}$ & $1.04097^{+0.00061}_{-0.00062}$ \\[0.4em]

$\tau$ & $0.056^{+0.015}_{-0.014}$ & $0.056^{+0.016}_{-0.014}$ & $0.056^{+0.015}_{-0.014}$ & $0.056^{+0.016}_{-0.015}$ \\[0.4em]

& $0.056^{+0.015}_{-0.014}$ & $0.055^{+0.015}_{-0.015}$ & $0.056^{+0.015}_{-0.014}$ & $0.056^{+0.015}_{-0.014}$ \\[0.4em]

${\rm{ln}}(10^{10} A_{\rm s})$ & $3.153^{+0.032}_{-0.031}$ & $3.154^{+0.033}_{-0.034}$ & $3.155^{+0.034}_{-0.033}$ & $3.153^{+0.033}_{-0.033}$ \\[0.4em]

& $3.047^{+0.029}_{-0.028}$ & $3.046^{+0.029}_{-0.028}$ & $3.046^{+0.029}_{-0.027}$ & $3.046^{+0.029}_{-0.028}$ \\[0.4em]

$H_0$ & $67.55^{+0.98}_{-1.0}$ & $67.5^{+1.1}_{-1.0}$ & $67.5^{+1.0}_{-1.0}$ & $67.6^{+1.1}_{-1.1}$ \\[0.4em]

& $67.60^{+0.99}_{-0.99}$ & $67.5^{+1.1}_{-1.1}$ & $67.5^{+1.1}_{-1.0}$ & $67.6^{+1.1}_{-1.1}$ \\[0.4em]

$S_8$ & $0.828^{+0.025}_{-0.024}$ & $0.829^{+0.025}_{-0.025}$ & $0.829^{+0.024}_{-0.025}$ & $0.827^{+0.026}_{-0.025}$ \\[0.4em]

& $0.827^{+0.024}_{-0.024}$ & $0.828^{+0.025}_{-0.025}$ & $0.829^{+0.025}_{-0.024}$ & $0.827^{+0.026}_{-0.026}$ \\[0.4em]

\noalign{\vskip 3pt}\hline\noalign{\vskip 1.5pt}\hline\noalign{\vskip 3pt}
\end{tabular}
\caption{\label{tab:table1_pantheon} Same as \tref{tab:table1_planck} but for Planck TTTEEE+lowE+lensing+Pantheon+ dataset.}
\end{table*}

\begin{table*}\centering
\begin{tabular} { l  c c c c}
\noalign{\vskip 3pt}\hline\noalign{\vskip 1.5pt}\hline\noalign{\vskip 5pt}
 &  \multicolumn{1}{c}{\bf\boldmath $R^2$ Model} &  \multicolumn{1}{c}{\bf\boldmath $R^3$ Model} &  \multicolumn{1}{c}{\bf Weyl Model} &  \multicolumn{1}{c}{\bf\boldmath $\Lambda$CDM}\\
\noalign{\vskip 3pt}\cline{2-5}\noalign{\vskip 3pt}

Parameter &  95\% limits &  95\% limits &  95\% limits &  95\% limits\\[0.4em]
\noalign{\vskip 3pt}\hline\noalign{\vskip 1.5pt}\hline\noalign{\vskip 3pt}
$\Omega_{\rm b} h^2$ & $0.02242^{+0.00026}_{-0.00026}$ & $0.02242^{+0.00026}_{-0.00026}$ & $0.02241^{+0.00027}_{-0.00027}$ & $0.02242^{+0.00027}_{-0.00026}$\\[0.4em]

& $0.02242^{+0.00026}_{-0.00026}$ & $0.02242^{+0.00026}_{-0.00026}$ & $0.02242^{+0.00026}_{-0.00026}$ & $0.02242^{+0.00027}_{-0.00027}$ \\[0.4em]

$\Omega_{\rm c} h^2$ & $0.1194^{+0.0018}_{-0.0017}$ & $0.1195^{+0.0019}_{-0.0018}$ & $0.1195^{+0.0018}_{-0.0019}$ & $0.1193^{+0.0018}_{-0.0018}$ \\[0.4em]

& $0.1194^{+0.0018}_{-0.0018}$ & $0.1194^{+0.0018}_{-0.0018}$ & $0.1194^{+0.0019}_{-0.0018}$ & $0.1193^{+0.0019}_{-0.0019}$\\[0.4em]

$100\theta_{\rm MC}$ & $1.04100^{+0.00056}_{-0.00058}$ & $1.04100^{+0.00057}_{-0.00057}$ & $1.04100^{+0.00058}_{-0.00057}$ & $1.04101^{+0.00058}_{-0.00058}$ \\[0.4em]

& $1.04100^{+0.00057}_{-0.00057}$ & $1.04100^{+0.00058}_{-0.00056}$ & $1.04099^{+0.00056}_{-0.00058}$ & $1.04100^{+0.00058}_{-0.00057}$ \\[0.4em]

$\tau$ & $0.057^{+0.014}_{-0.014}$ & $0.057^{+0.015}_{-0.014}$ & $0.057^{+0.015}_{-0.014}$ & $0.057^{+0.015}_{-0.014}$ \\[0.4em]

& $0.058^{+0.015}_{-0.014}$ & $0.057^{+0.014}_{-0.014}$ & $0.057^{+0.015}_{-0.014}$ & $0.057^{+0.015}_{-0.014}$ \\[0.4em]

${\rm{ln}}(10^{10} A_{\rm s})$ & $3.155^{+0.031}_{-0.030}$ & $3.156^{+0.033}_{-0.033}$ & $3.156^{+0.033}_{-0.033}$ & $3.155^{+0.032}_{-0.032}$ \\[0.4em]

& $3.050^{+0.028}_{-0.027}$ & $3.050^{+0.028}_{-0.027}$ & $3.050^{+0.029}_{-0.027}$ & $3.049^{+0.029}_{-0.028}$ \\[0.4em]

$H_0$ & $67.63^{+0.79}_{-0.81}$ & $67.61^{+0.83}_{-0.82}$ & $67.60^{+0.86}_{-0.81}$ & $67.66^{+0.84}_{-0.81}$ \\[0.4em]

& $67.65^{+0.80}_{-0.82}$ & $67.62^{+0.83}_{-0.82}$ & $67.62^{+0.84}_{-0.82}$ & $67.66^{+0.85}_{-0.83}$ \\[0.4em]

$S_8$ & $0.827^{+0.020}_{-0.020}$ & $0.828^{+0.020}_{-0.020}$ & $0.828^{+0.020}_{-0.020}$ & $0.826^{+0.020}_{-0.020}$ \\[0.4em]

& $0.827^{+0.020}_{-0.020}$ & $0.828^{+0.020}_{-0.020}$ & $0.827^{+0.020}_{-0.020}$ & $0.826^{+0.021}_{-0.021}$ \\[0.4em]

\noalign{\vskip 3pt}\hline\noalign{\vskip 1.5pt}\hline\noalign{\vskip 3pt}
\end{tabular}

\caption{\label{tab:table1_bao_bk18} Same as \tref{tab:table1_planck} but for Planck TTTEEE+lowE+lensing+BAO+BK18 dataset.}
\end{table*}

\begin{table*}\centering
\begin{tabular} { l  c c c c}
\noalign{\vskip 3pt}\hline\noalign{\vskip 1.5pt}\hline\noalign{\vskip 5pt}
 &  \multicolumn{1}{c}{\bf\boldmath $R^2$ Model} &  \multicolumn{1}{c}{\bf\boldmath $R^3$ Model} &  \multicolumn{1}{c}{\bf Weyl Model} &  \multicolumn{1}{c}{\bf\boldmath $\Lambda$CDM}\\
\noalign{\vskip 3pt}\cline{2-5}\noalign{\vskip 3pt}

 Parameter &  95\% limits &  95\% limits &  95\% limits &  95\% limits\\[0.4em]
\noalign{\vskip 3pt}\hline\noalign{\vskip 1.5pt}\hline\noalign{\vskip 3pt}
$\Omega_{\rm b} h^2$ & $0.02249^{+0.00025}_{-0.00025}$ & $0.02250^{+0.00026}_{-0.00026}$ & $0.02250^{+0.00026}_{-0.00026}$ & $0.02249^{+0.00026}_{-0.00026}$ \\[0.4em]

& $0.02252^{+0.00026}_{-0.00025}$ & $0.02252^{+0.00026}_{-0.00026}$ & $0.02249^{+0.00025}_{-0.00026}$ & $0.02249^{+0.00026}_{-0.00026}$ \\[0.4em]

$\Omega_{\rm c} h^2$ & $0.1185^{+0.0016}_{-0.0015}$ & $0.1185^{+0.0017}_{-0.0017}$ & $0.1185^{+0.0017}_{-0.0017}$ & $0.1184^{+0.0017}_{-0.0017}$ \\[0.4em]

& $0.1182^{+0.0016}_{-0.0015}$ & $0.1182^{+0.0017}_{-0.0017}$ & $0.1185^{+0.0017}_{-0.0017}$ & $0.1184^{+0.0017}_{-0.0017}$ \\[0.4em]

$100\theta_{\rm MC}$ & $1.04107^{+0.00057}_{-0.00058}$ & $1.04108^{+0.00056}_{-0.00058}$ & $1.04108^{+0.00056}_{-0.00057}$ & $1.04108^{+0.00055}_{-0.00057}$\\[0.4em]

& $1.04109^{+0.00056}_{-0.00057}$ & $1.04110^{+0.00056}_{-0.00057}$ & $1.04107^{+0.00058}_{-0.00056}$ & $1.04108^{+0.00057}_{-0.00056}$ \\[0.4em]

$\tau$ & $0.057^{+0.014}_{-0.014}$ & $0.057^{+0.015}_{-0.014}$ & $0.057^{+0.015}_{-0.014}$ & $0.057^{+0.015}_{-0.014}$ \\[0.4em]

& $0.057^{+0.015}_{-0.014}$ & $0.057^{+0.015}_{-0.014}$ & $0.057^{+0.016}_{-0.014}$ & $0.057^{+0.015}_{-0.014}$ \\[0.4em]

${\rm{ln}}(10^{10} A_{\rm s})$ & $3.147^{+0.031}_{-0.030}$ & $3.147^{+0.032}_{-0.031}$ & $3.148^{+0.033}_{-0.031}$ & $3.146^{+0.033}_{-0.033}$ \\[0.4em]

& $3.046^{+0.029}_{-0.027}$ & $3.046^{+0.029}_{-0.028}$ & $3.047^{+0.030}_{-0.028}$ & $3.047^{+0.029}_{-0.028}$ \\[0.4em]

$H_0$ & $68.02^{+0.71}_{-0.74}$ & $68.04^{+0.76}_{-0.76}$ & $68.04^{+0.77}_{-0.76}$ & $68.05^{+0.78}_{-0.77}$ \\[0.4em]

& $68.19^{+0.72}_{-0.72}$ & $68.18^{+0.77}_{-0.76}$ & $68.02^{+0.76}_{-0.76}$ & $68.07^{+0.79}_{-0.76}$ \\[0.4em]

$S_8$ & $0.816^{+0.018}_{-0.018}$ & $0.816^{+0.018}_{-0.018}$ & $0.816^{+0.018}_{-0.018}$ & $0.815^{+0.019}_{-0.018}$ \\[0.4em]

& $0.812^{+0.018}_{-0.018}$ & $0.812^{+0.018}_{-0.018}$ & $0.816^{+0.019}_{-0.018}$ & $0.815^{+0.019}_{-0.018}$ \\[0.4em]

\noalign{\vskip 3pt}\hline\noalign{\vskip 1.5pt}\hline\noalign{\vskip 3pt}
\end{tabular}
\caption{\label{tab:table1_bao_des} Same as \tref{tab:table1_planck} but for Planck TTTEEE+lowE+lensing+BAO+DES dataset.}
\end{table*}

\begin{figure*}[h!]
\includegraphics[width=\textwidth]{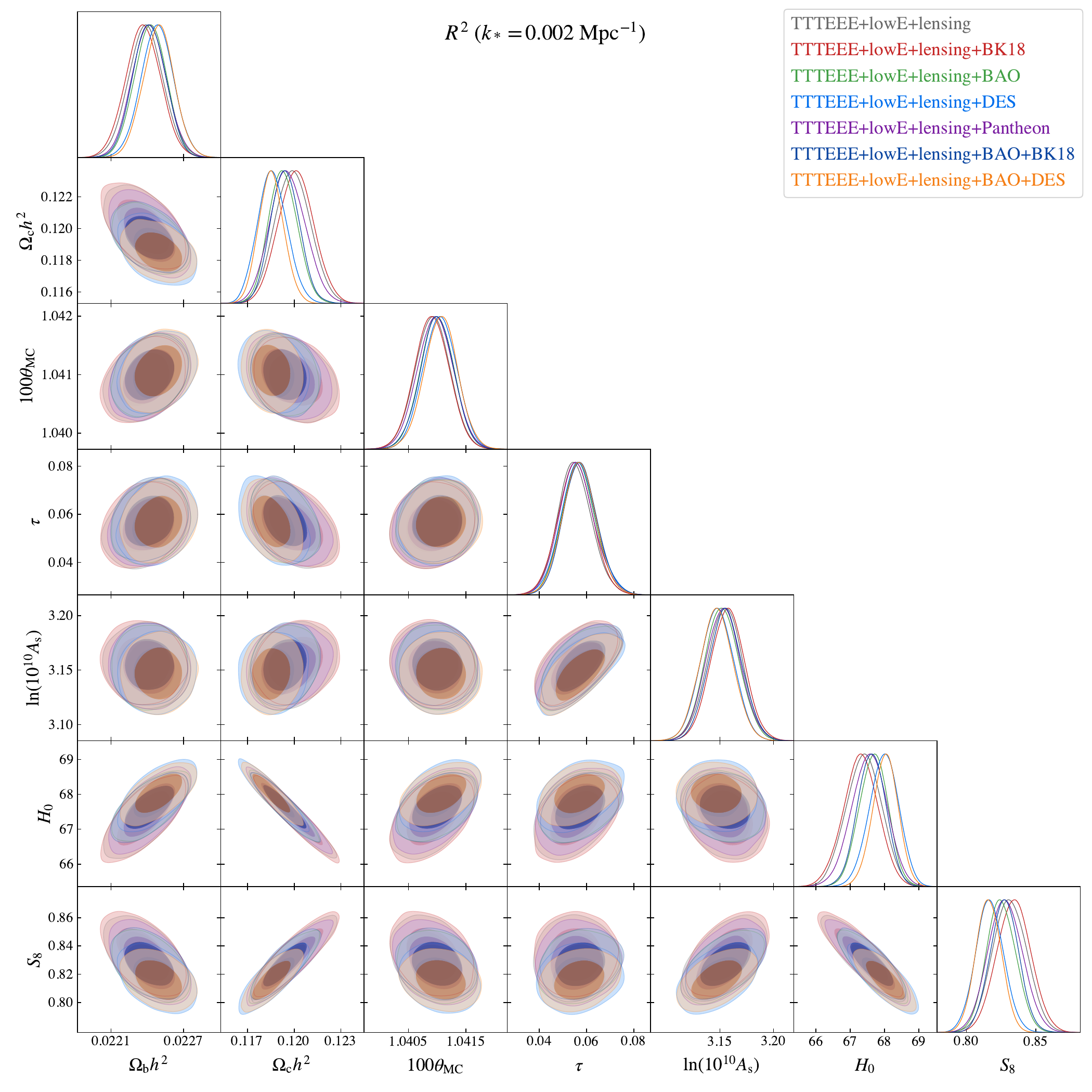}
\caption{\label{fig:all_data_star_plot1} Marginalized 68\% and 95\% contour plots for the main cosmological parameters for $R^2$ Starobinsky model for each dataset in \tref{tab:datacomb} for $k_* = 0.002\ {\rm Mpc}^{-1}$.}
\end{figure*}

\newpage
${}$

\begin{figure*}[h!]
\includegraphics[width=\textwidth]{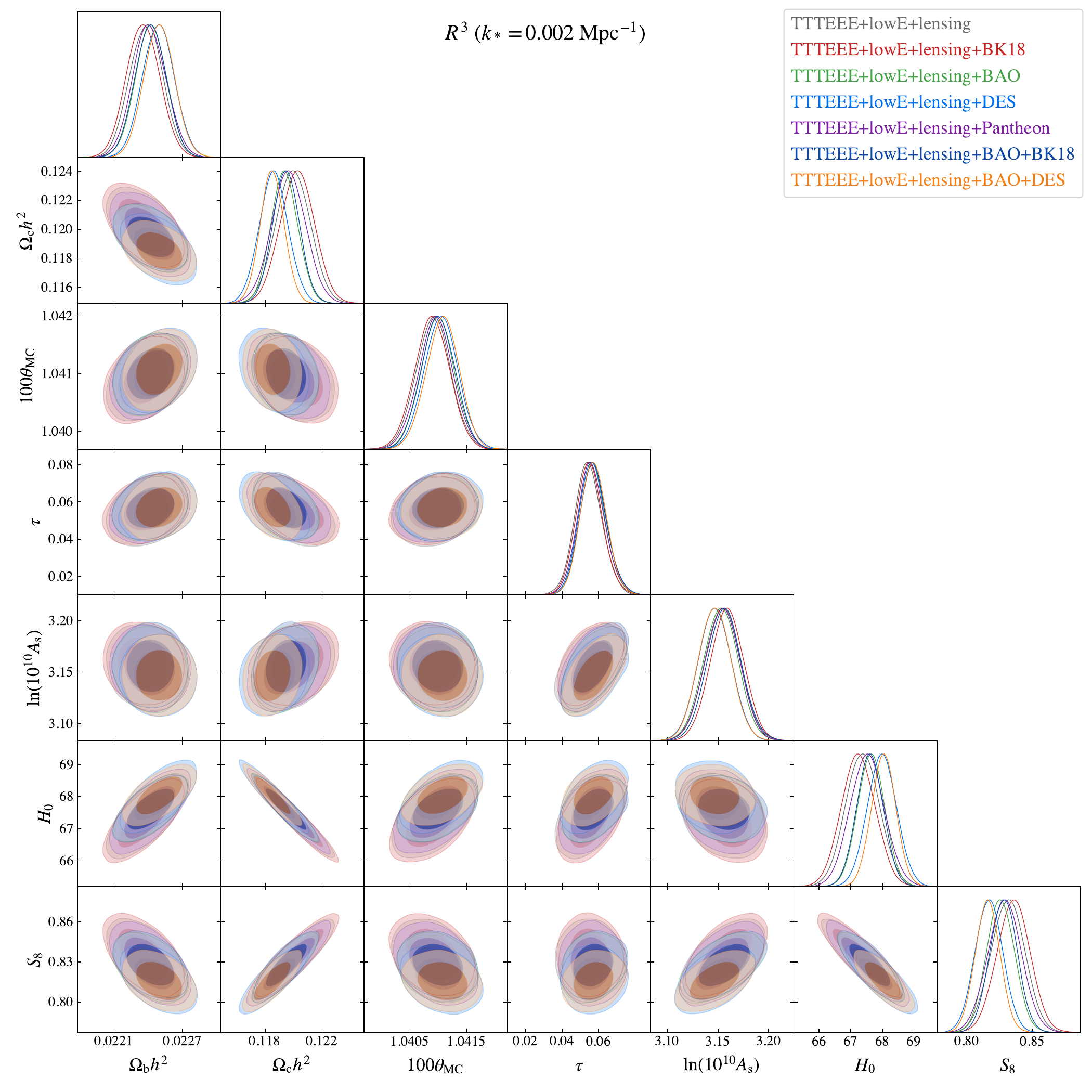}
\caption{\label{fig:all_data_exst_plot1} Same as \fref{fig:all_data_star_plot1} but for $R^3$ Starobinsky model.}
\end{figure*}

\newpage
${}$

\begin{figure*}[h!]
\includegraphics[width=\textwidth]{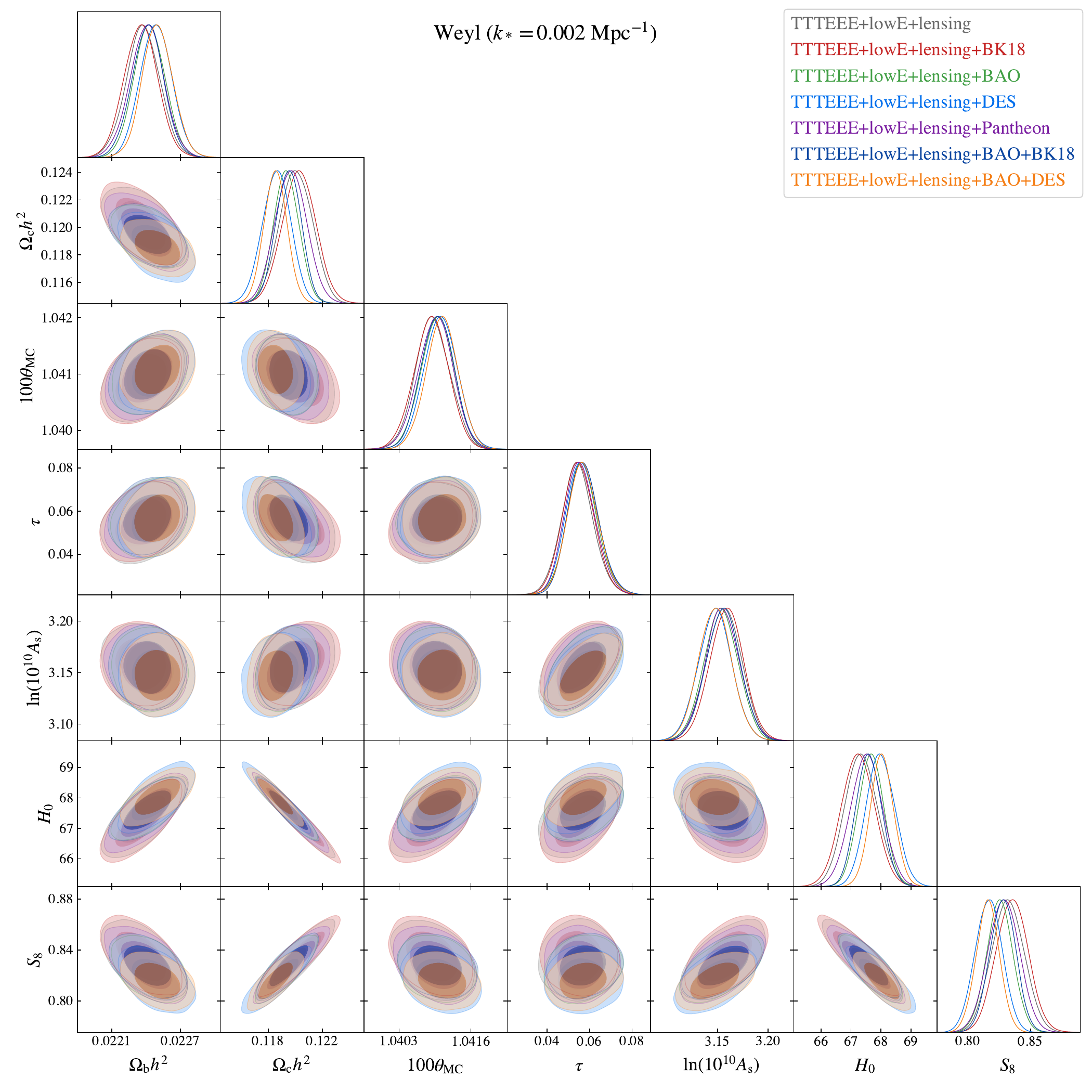}
\caption{\label{fig:all_data_weyl_plot1} Same as \fref{fig:all_data_star_plot1} but for Weyl model.}
\end{figure*}

\newpage
${}$

\begin{figure*}[h!]
\includegraphics[width=\textwidth]{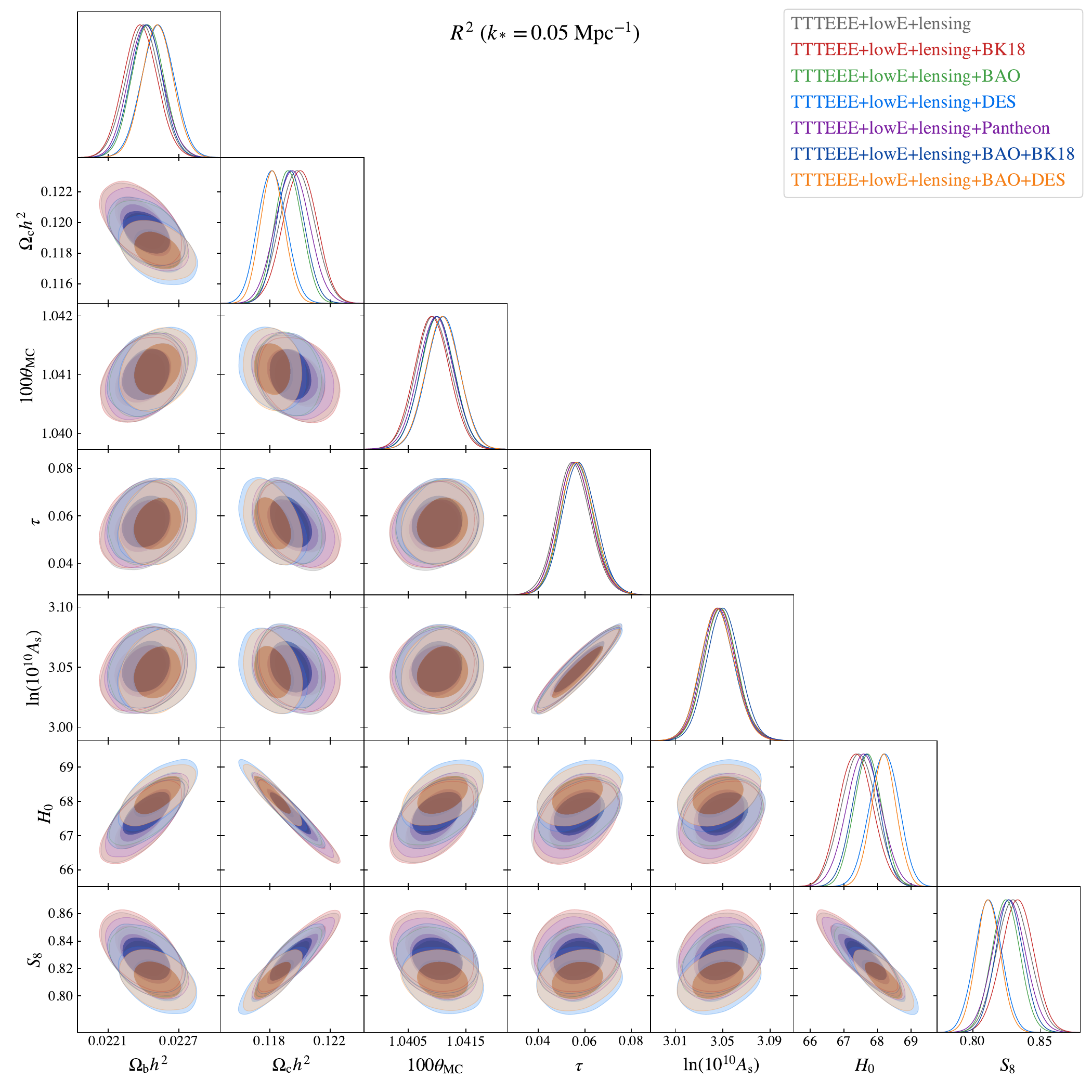}
\caption{\label{fig:all_data_star5_plot1} Marginalized 68\% and 95\% contour plots for the main cosmological parameters for $R^2$ Starobinsky model for each dataset in \tref{tab:datacomb} for $k_* = 0.05\ {\rm Mpc}^{-1}$.}
\end{figure*}

\newpage
${}$

\begin{figure*}[h!]
\includegraphics[width=\textwidth]{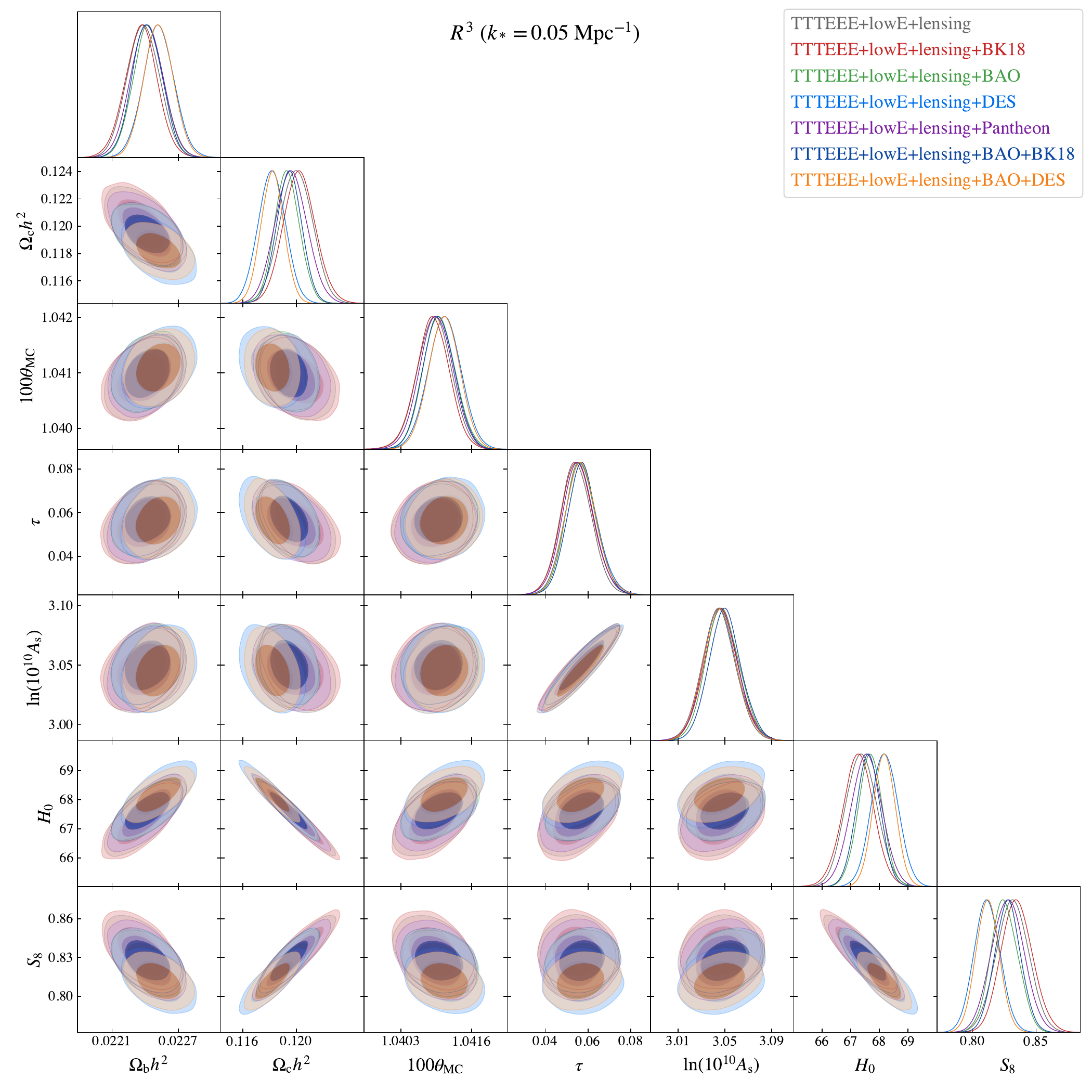}
\caption{\label{fig:all_data_exst5_plot1} Same as \fref{fig:all_data_star5_plot1} but for $R^3$ Starobinsky model.}
\end{figure*}

\newpage
${}$

\begin{figure*}[h!]
\includegraphics[width=\textwidth]{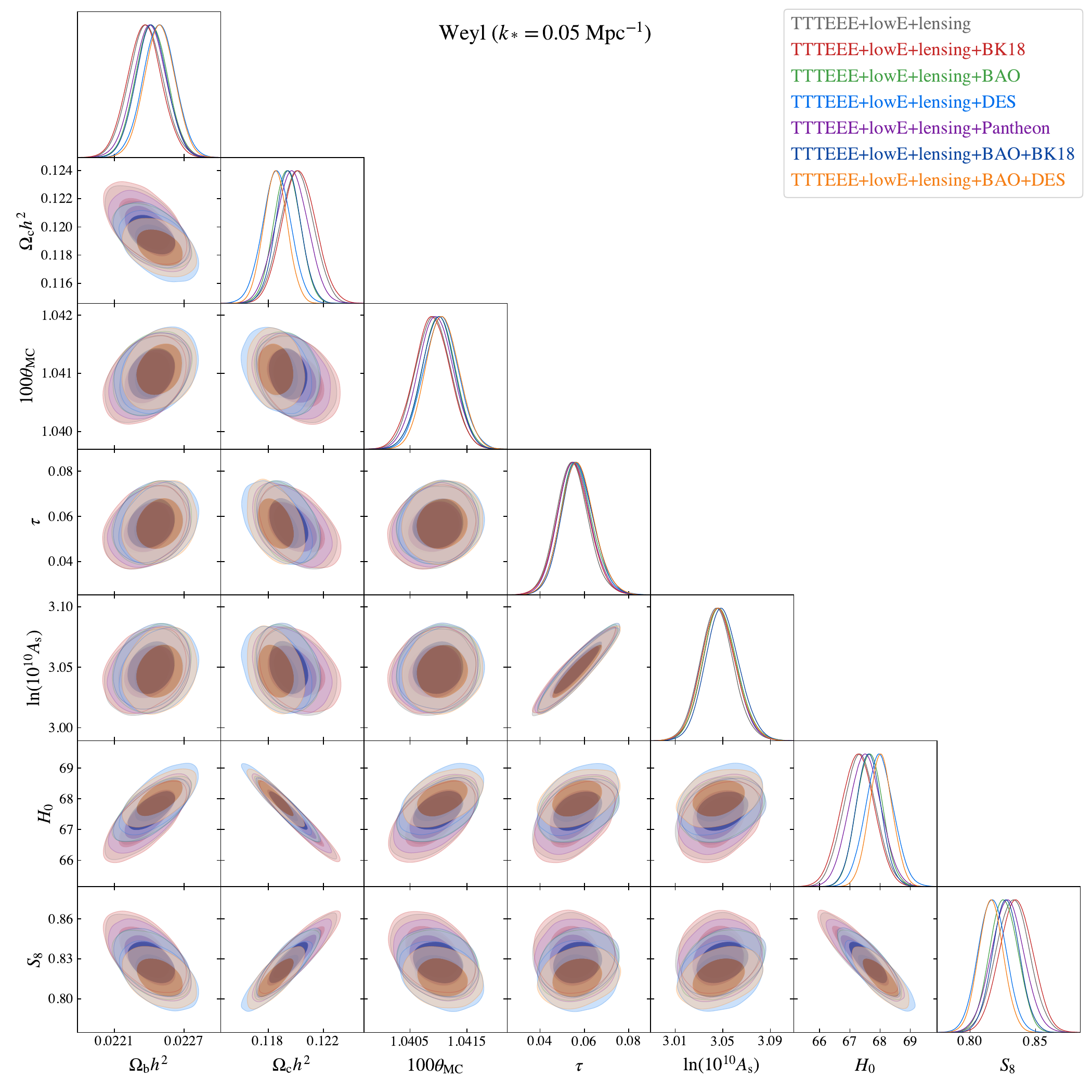}
\caption{\label{fig:all_data_weyl5_plot1} Same as \fref{fig:all_data_star5_plot1} but for Weyl model.}
\end{figure*}

\newpage
${}$
\newpage
${}$
\newpage
${}$
\newpage
${}$
\newpage
${}$

\subsubsection{Power spectrum parameters}
\label{ssec:ps_params}

We summarize the constraints on the inflationary parameters in this section with $k_* = 0.002, 0.05\ {\rm Mpc}^{-1}$. The parameters are $n_{\rm s}$, $r$, $n_{\rm run}$, $n_{\rm run,run}$, $n_{\rm t}$ and $n_{\rm t, run}$

\begin{table*}[h!]\centering
\begin{tabular} { l  c c c c}
\noalign{\vskip 3pt}\hline\noalign{\vskip 1.5pt}\hline\noalign{\vskip 5pt}
 \multicolumn{1}{c}{\bf } &  \multicolumn{1}{c}{\bf\boldmath $R^2$ Model} &  \multicolumn{1}{c}{\bf\boldmath $R^3$ Model} &  \multicolumn{1}{c}{\bf Weyl Model} &  \multicolumn{1}{c}{\bf\boldmath $\Lambda$CDM}\\
\noalign{\vskip 3pt}\cline{2-5}\noalign{\vskip 3pt}

 Parameter &  95\% limits &  95\% limits &  95\% limits &  95\% limits\\[0.4em]
\noalign{\vskip 3pt}\hline\noalign{\vskip 1.5pt}\hline\noalign{\vskip 3pt}

$n_{\rm s}$ & $0.9671^{+0.0063}_{-0.0068}$ & $0.9668^{+0.0088}_{-0.0086}$ & $0.9661^{+0.0089}_{-0.0082}$ & $0.9660^{+0.0080}_{-0.0084}$ \\[0.4em]

& $0.9660^{+0.0070}_{-0.0068}$ & $0.9651^{+0.0081}_{-0.0080}$ & $0.9648^{+0.0080}_{-0.0079}$ & $0.9658^{+0.0084}_{-0.0083}$ \\[0.4em]

$r$ & $0.0031^{+0.0012}_{-0.0011}$ & $0.0034^{+0.0027}_{-0.0023}$ & $0.0026^{+0.0021}_{-0.0017}$ & $< 0.115$ \\[0.4em]

& $0.0033^{+0.0014}_{-0.0012}$ & $0.0031^{+0.0026}_{-0.0022}$ & $0.0024^{+0.0020}_{-0.0015}$ & $< 0.117$ \\[0.4em]

$n_{\rm run}$ & $-0.00055^{+0.00020}_{-0.00023}$ & $-0.00060^{+0.00040}_{-0.00048}$ & $-0.00045^{+0.00030}_{-0.00037}$ & -- \\[0.4em]

& $-0.00059^{+0.00022}_{-0.00025}$ & $-0.00056^{+0.00038}_{-0.00045}$ & $-0.00043^{+0.00027}_{-0.00036}$ & -- \\[0.4em]

$n_{\rm run,run}$ & $\left(\,1.9^{+1.2}_{-1.0}\,\right)\cdot 10^{-5}$ & $\left(\,2.0^{+1.5}_{-1.2}\,\right)\cdot 10^{-5}$ & $\left(\,1.52^{+1.2}_{-0.88}\,\right)\cdot 10^{-5}$ & -- \\[0.4em]

& $\left(\,2.1^{+1.3}_{-1.1}\,\right)\cdot 10^{-5}$ & $\left(\,2.0^{+1.7}_{-1.2}\,\right)\cdot 10^{-5}$ & $\left(\,1.51^{+1.2}_{-0.87}\,\right)\cdot 10^{-5}$ & -- \\[0.4em]

$n_{\rm t}$ & $-0.00039^{+0.00014}_{-0.00015}$ & $-0.00042^{+0.00029}_{-0.00034}$ & $-0.00032^{+0.00021}_{-0.00027}$ & -- \\[0.4em]

& $-0.00042^{+0.00015}_{-0.00017}$ & $-0.00039^{+0.00027}_{-0.00032}$ & $-0.00030^{+0.00019}_{-0.00025}$ & -- \\[0.4em]

$n_{\rm t, run}$ & $\left(\,-13.0^{+6.7}_{-7.9}\,\right)\cdot 10^{-6}$ & $\left(\,-14.0^{+7.9}_{-10}\,\right)\cdot 10^{-6}$ & $\left(\,-10.4^{+6.0}_{-7.8}\,\right)\cdot 10^{-6}$ & -- \\[0.4em]

& $\left(\,-14.3^{+7.5}_{-8.6}\,\right)\cdot 10^{-6}$ & $\left(\,-13.8^{+7.7}_{-11}\,\right)\cdot 10^{-6}$ & $\left(\,-10.4^{+5.9}_{-8.0}\,\right)\cdot 10^{-6}$ & -- \\[0.4em]

\noalign{\vskip 3pt}\hline\noalign{\vskip 1.5pt}\hline\noalign{\vskip 3pt}
\end{tabular}
\caption{\label{tab:table2_planck} 95\% Confidence limits of the power spectrum parameters for Planck TTTEEE+lowE+lensing dataset.  The upper set of figures for each parameter are constraints at $k_* = 0.002\ {\rm Mpc}^{-1}$ and the lower set of figures are constraints at $k_* = 0.05\ {\rm Mpc}^{-1}$.}
\end{table*}

\begin{table*}\centering
\begin{tabular} { l  c c c c}
\noalign{\vskip 3pt}\hline\noalign{\vskip 1.5pt}\hline\noalign{\vskip 5pt}
 \multicolumn{1}{c}{\bf } &  \multicolumn{1}{c}{\bf\boldmath $R^2$ Model} &  \multicolumn{1}{c}{\bf\boldmath $R^3$ Model} &  \multicolumn{1}{c}{\bf Weyl Model} &  \multicolumn{1}{c}{\bf\boldmath $\Lambda$CDM}\\
\noalign{\vskip 3pt}\cline{2-5}\noalign{\vskip 3pt}

 Parameter &  95\% limits &  95\% limits &  95\% limits &  95\% limits\\[0.4em]
\noalign{\vskip 3pt}\hline\noalign{\vskip 1.5pt}\hline\noalign{\vskip 3pt}

$n_{\rm s}$ & $0.9678^{+0.0060}_{-0.0063}$ & $0.9677^{+0.0087}_{-0.0084}$ & $0.9670^{+0.0087}_{-0.0083}$ & $0.9671^{+0.0083}_{-0.0085}$ \\[0.4em]

& $0.9666^{+0.0067}_{-0.0070}$ & $0.9660^{+0.0079}_{-0.0080}$ & $0.9656^{+0.0080}_{-0.0078}$ & $0.9668^{+0.0084}_{-0.0081}$ \\[0.4em]

$r$ & $0.0030^{+0.0012}_{-0.0010}$ & $0.0034^{+0.0027}_{-0.0023}$ & $0.0026^{+0.0022}_{-0.0017}$ & $< 0.113$ \\[0.4em]

& $0.0032^{+0.0013}_{-0.0012}$ & $0.0032^{+0.0027}_{-0.0022}$ & $0.0025^{+0.0020}_{-0.0016}$ & $< 0.125$ \\[0.4em]

$n_{\rm run}$ & $-0.00053^{+0.00019}_{-0.00021}$ & $-0.00060^{+0.00039}_{-0.00047}$ & $-0.00047^{+0.00030}_{-0.00039}$ & -- \\[0.4em]

& $-0.00057^{+0.00022}_{-0.00024}$ & $-0.00056^{+0.00038}_{-0.00047}$ & $-0.00044^{+0.00028}_{-0.00036}$ & -- \\[0.4em]

$n_{\rm run,run}$ & $\left(\,1.76^{+1.1}_{-0.90}\,\right)\cdot 10^{-5}$ & $\left(\,2.0^{+1.5}_{-1.1}\,\right)\cdot 10^{-5}$ & $\left(\,1.53^{+1.2}_{-0.86}\,\right)\cdot 10^{-5}$ & -- \\[0.4em]

& $\left(\,2.0^{+1.3}_{-1.1}\,\right)\cdot 10^{-5}$ & $\left(\,2.0^{+1.7}_{-1.1}\,\right)\cdot 10^{-5}$ & $\left(\,1.53^{+1.2}_{-0.87}\,\right)\cdot 10^{-5}$ & -- \\[0.4em]

$n_{\rm t}$ & $-0.00037^{+0.00013}_{-0.00015}$ & $-0.00042^{+0.00028}_{-0.00034}$ & $-0.00033^{+0.00022}_{-0.00028}$ & -- \\[0.4em]

& $-0.00040^{+0.00015}_{-0.00016}$ & $-0.00040^{+0.00027}_{-0.00034}$ & $-0.00031^{+0.00020}_{-0.00025}$ & -- \\[0.4em]

$n_{\rm t, run}$ & $\left(\,-12.1^{+6.1}_{-7.3}\,\right)\cdot 10^{-6}$ & $\left(\,-13.6^{+7.6}_{-10}\,\right)\cdot 10^{-6}$ & $\left(\,-10.5^{+5.9}_{-7.7}\,\right)\cdot 10^{-6}$ & -- \\[0.4em]

& $\left(\,-13.6^{+7.2}_{-8.4}\,\right)\cdot 10^{-6}$ & $\left(\,-13.6^{+7.6}_{-11}\,\right)\cdot 10^{-6}$ & $\left(\,-10.5^{+5.9}_{-7.9}\,\right)\cdot 10^{-6}$ & -- \\[0.4em]

\noalign{\vskip 3pt}\hline\noalign{\vskip 1.5pt}\hline\noalign{\vskip 3pt}
\end{tabular}
\caption{\label{tab:table2_pantheon} Same as \tref{tab:table2_planck} but for Planck TTTEEE+lowE+lensing+Pantheon+ dataset.}
\end{table*}

\begin{table*}\centering
\begin{tabular} { l  c c c c}
\noalign{\vskip 3pt}\hline\noalign{\vskip 1.5pt}\hline\noalign{\vskip 5pt}
 \multicolumn{1}{c}{\bf } &  \multicolumn{1}{c}{\bf\boldmath $R^2$ Model} &  \multicolumn{1}{c}{\bf\boldmath $R^3$ Model} &  \multicolumn{1}{c}{\bf Weyl Model} &  \multicolumn{1}{c}{\bf\boldmath $\Lambda$CDM}\\
\noalign{\vskip 3pt}\cline{2-5}\noalign{\vskip 3pt}

 Parameter &  95\% limits &  95\% limits &  95\% limits &  95\% limits\\[0.4em]
\noalign{\vskip 3pt}\hline\noalign{\vskip 1.5pt}\hline\noalign{\vskip 3pt}

$n_{\rm s}$ & $0.9682^{+0.0056}_{-0.0060}$ & $0.9680^{+0.0077}_{-0.0078}$ & $0.9679^{+0.0079}_{-0.0077}$ & $0.9668^{+0.0072}_{-0.0072}$ \\[0.4em]

& $0.9669^{+0.0063}_{-0.0068}$ & $0.9663^{+0.0076}_{-0.0074}$ & $0.9663^{+0.0072}_{-0.0071}$ & $0.9668^{+0.0073}_{-0.0073}$ \\[0.4em]

$r$ & $0.00293^{+0.0011}_{-0.00098}$ & $0.0034^{+0.0027}_{-0.0023}$ & $0.0029^{+0.0022}_{-0.0018}$ & $< 0.0313$ \\[0.4em]

& $0.0032^{+0.0012}_{-0.0011}$ & $0.0032^{+0.0026}_{-0.0021}$ & $0.0027^{+0.0020}_{-0.0017}$ & $< 0.0358$ \\[0.4em]

$n_{\rm run}$ & $-0.00052^{+0.00018}_{-0.00020}$ & $-0.00060^{+0.00040}_{-0.00048}$ & $-0.00050^{+0.00032}_{-0.00039}$ & -- \\[0.4em]

& $-0.00056^{+0.00021}_{-0.00023}$ & $-0.00057^{+0.00037}_{-0.00045}$ & $-0.00047^{+0.00029}_{-0.00036}$ & -- \\[0.4em]

$n_{\rm run,run}$ & $\left(\,1.71^{+1.0}_{-0.84}\,\right)\cdot 10^{-5}$ & $\left(\,2.0^{+1.7}_{-1.1}\,\right)\cdot 10^{-5}$ & $\left(\,1.61^{+1.1}_{-0.91}\,\right)\cdot 10^{-5}$ & -- \\[0.4em]

& $\left(\,1.9^{+1.2}_{-1.0}\,\right)\cdot 10^{-5}$ & $\left(\,2.0^{+1.7}_{-1.1}\,\right)\cdot 10^{-5}$ & $\left(\,1.60^{+1.2}_{-0.91}\,\right)\cdot 10^{-5}$ & -- \\[0.4em]

$n_{\rm t}$ & $-0.00037^{+0.00012}_{-0.00014}$ & $-0.00043^{+0.00028}_{-0.00034}$ & $-0.00036^{+0.00023}_{-0.00028}$ & -- \\[0.4em]

& $-0.00040^{+0.00014}_{-0.00016}$ & $-0.00040^{+0.00027}_{-0.00032}$ & $-0.00033^{+0.00021}_{-0.00025}$ & -- \\[0.4em]

$n_{\rm t, run}$ & $\left(\,-11.7^{+5.6}_{-6.9}\,\right)\cdot 10^{-6}$ & $\left(\,-13.7^{+7.6}_{-11}\,\right)\cdot 10^{-6}$ & $\left(\,-11.0^{+6.2}_{-7.3}\,\right)\cdot 10^{-6}$ & -- \\[0.4em]

& $\left(\,-13.2^{+6.8}_{-8.0}\,\right)\cdot 10^{-6}$ & $\left(\,-13.6^{+7.6}_{-11}\,\right)\cdot 10^{-6}$ & $\left(\,-11.0^{+6.1}_{-7.7}\,\right)\cdot 10^{-6}$ & -- \\[0.4em]

\noalign{\vskip 3pt}\hline\noalign{\vskip 1.5pt}\hline\noalign{\vskip 3pt}
\end{tabular}
\caption{\label{tab:table2_bao_bk18} Same as \tref{tab:table2_planck} but for Planck TTTEEE+lowE+lensing+BAO+BK18 dataset.}
\end{table*}

\begin{table*}\centering
\begin{tabular} { l  c c c c}
\noalign{\vskip 3pt}\hline\noalign{\vskip 1.5pt}\hline\noalign{\vskip 5pt}
 \multicolumn{1}{c}{\bf } &  \multicolumn{1}{c}{\bf\boldmath $R^2$ Model} &  \multicolumn{1}{c}{\bf\boldmath $R^3$ Model} &  \multicolumn{1}{c}{\bf Weyl Model} &  \multicolumn{1}{c}{\bf\boldmath $\Lambda$CDM}\\
\noalign{\vskip 3pt}\cline{2-5}\noalign{\vskip 3pt}

 Parameter &  95\% limits &  95\% limits &  95\% limits &  95\% limits\\[0.4em]
\noalign{\vskip 3pt}\hline\noalign{\vskip 1.5pt}\hline\noalign{\vskip 3pt}

$n_{\rm s}$ & $0.9696^{+0.0047}_{-0.0054}$ & $0.9700^{+0.0075}_{-0.0076}$ & $0.9695^{+0.0075}_{-0.0074}$ & $0.9690^{+0.0074}_{-0.0072}$ \\[0.4em]

& $0.9688^{+0.0054}_{-0.0058}$ & $0.9686^{+0.0071}_{-0.0071}$ & $0.9680^{+0.0070}_{-0.0071}$ & $0.9691^{+0.0073}_{-0.0073}$ \\[0.4em]

$r$ & $0.00268^{+0.00097}_{-0.00078}$ & $0.0035^{+0.0028}_{-0.0023}$ & $0.0028^{+0.0024}_{-0.0017}$ & $< 0.114$ \\[0.4em]

& $0.00281^{+0.0011}_{-0.00090}$ & $0.0034^{+0.0026}_{-0.0021}$ & $0.0027^{+0.0021}_{-0.0016}$ & $< 0.132$ \\[0.4em]

$n_{\rm run}$ & $-0.00047^{+0.00014}_{-0.00018}$ & $-0.00062^{+0.00039}_{-0.00050}$ & $-0.00049^{+0.00030}_{-0.00041}$ & -- \\[0.4em]

& $-0.00050^{+0.00016}_{-0.00019}$ & $-0.00060^{+0.00037}_{-0.00046}$ & $-0.00047^{+0.00028}_{-0.00037}$ & -- \\[0.4em]

$n_{\rm run,run}$ & $\left(\,1.49^{+0.86}_{-0.64}\,\right)\cdot 10^{-5}$ & $\left(\,2.0^{+1.7}_{-1.1}\,\right)\cdot 10^{-5}$ & $\left(\,1.49^{+1.1}_{-0.82}\,\right)\cdot 10^{-5}$ & -- \\[0.4em]

& $\left(\,1.60^{+0.97}_{-0.74}\,\right)\cdot 10^{-5}$ & $\left(\,2.0^{+1.7}_{-1.1}\,\right)\cdot 10^{-5}$ & $\left(\,1.53^{+1.2}_{-0.84}\,\right)\cdot 10^{-5}$ & -- \\[0.4em]

$n_{\rm t}$ & $-0.000335^{+0.000098}_{-0.00012}$ & $-0.00044^{+0.00028}_{-0.00035}$ & $-0.00035^{+0.00022}_{-0.00029}$ & -- \\[0.4em]

& $-0.00035^{+0.00011}_{-0.00013}$ & $-0.00043^{+0.00026}_{-0.00032}$ & $-0.00034^{+0.00020}_{-0.00027}$ & -- \\[0.4em]

$n_{\rm t, run}$ & $\left(\,-10.3^{+4.3}_{-5.8}\,\right)\cdot 10^{-6}$ & $\left(\,-13.4^{+7.4}_{-11}\,\right)\cdot 10^{-6}$ & $\left(\,-10.3^{+5.5}_{-7.5}\,\right)\cdot 10^{-6}$ & -- \\[0.4em]

& $\left(\,-11.0^{+5.0}_{-6.5}\,\right)\cdot 10^{-6}$ & $\left(\,-13.5^{+7.5}_{-11}\,\right)\cdot 10^{-6}$ & $\left(\,-10.5^{+5.7}_{-7.7}\,\right)\cdot 10^{-6}$ & -- \\[0.4em]

\noalign{\vskip 3pt}\hline\noalign{\vskip 1.5pt}\hline\noalign{\vskip 3pt}
\end{tabular}
\caption{\label{tab:table2_bao_des} Same as \tref{tab:table2_planck} but for Planck TTTEEE+lowE+lensing+BAO+DES dataset.}
\end{table*}

\newpage
${}$
\newpage
${}$
\newpage
${}$

\begin{figure*}
\includegraphics[width=\textwidth]{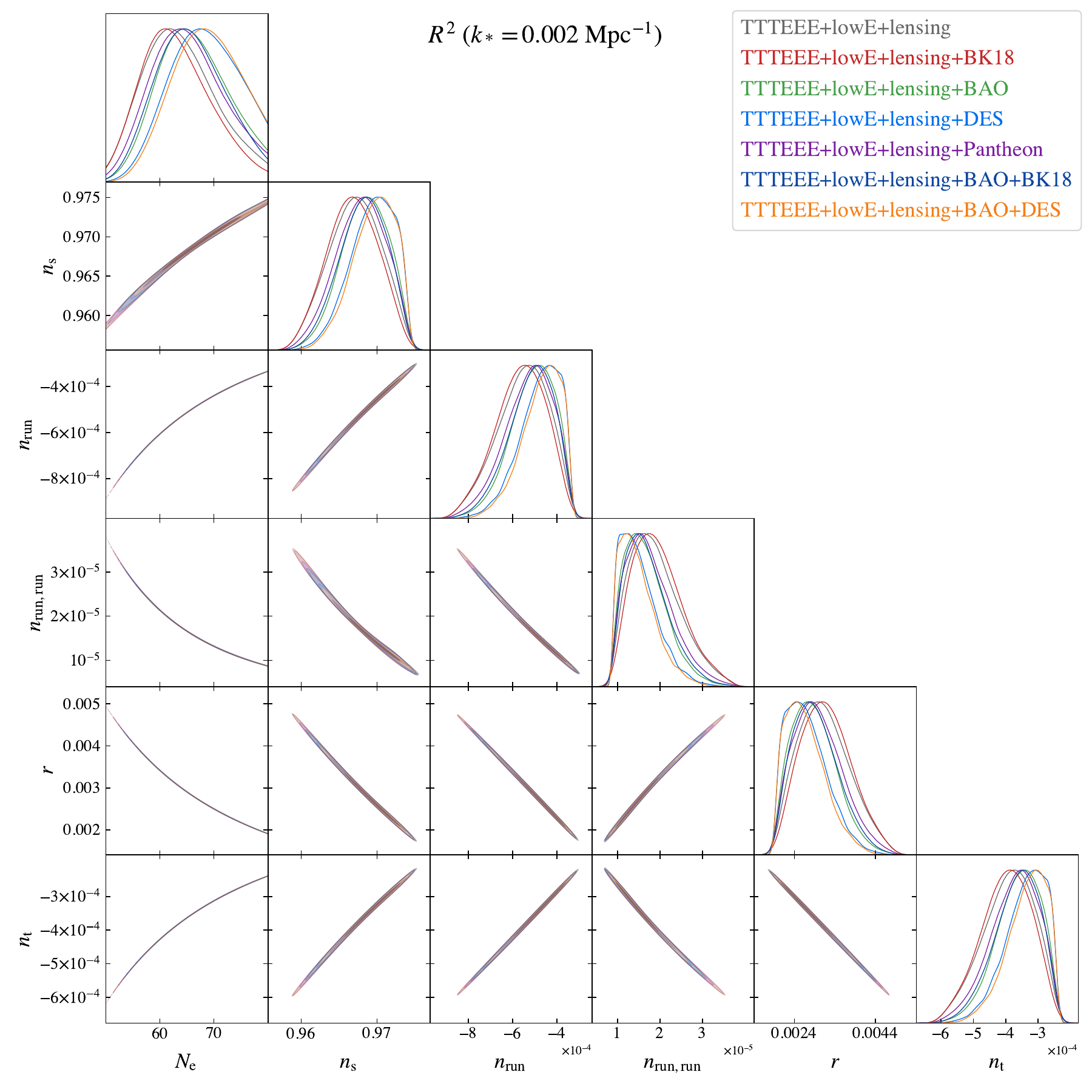}
\caption{\label{fig:all_data_star_plot2} Marginalized 68\% and 95\% contour plots for the power spectrum parameters for $R^2$ Starobinsky model for each dataset in \tref{tab:datacomb} for $k_* = 0.002\ {\rm Mpc}^{-1}$.}
\end{figure*}

\newpage
${}$

\begin{figure*}
\includegraphics[width=\textwidth]{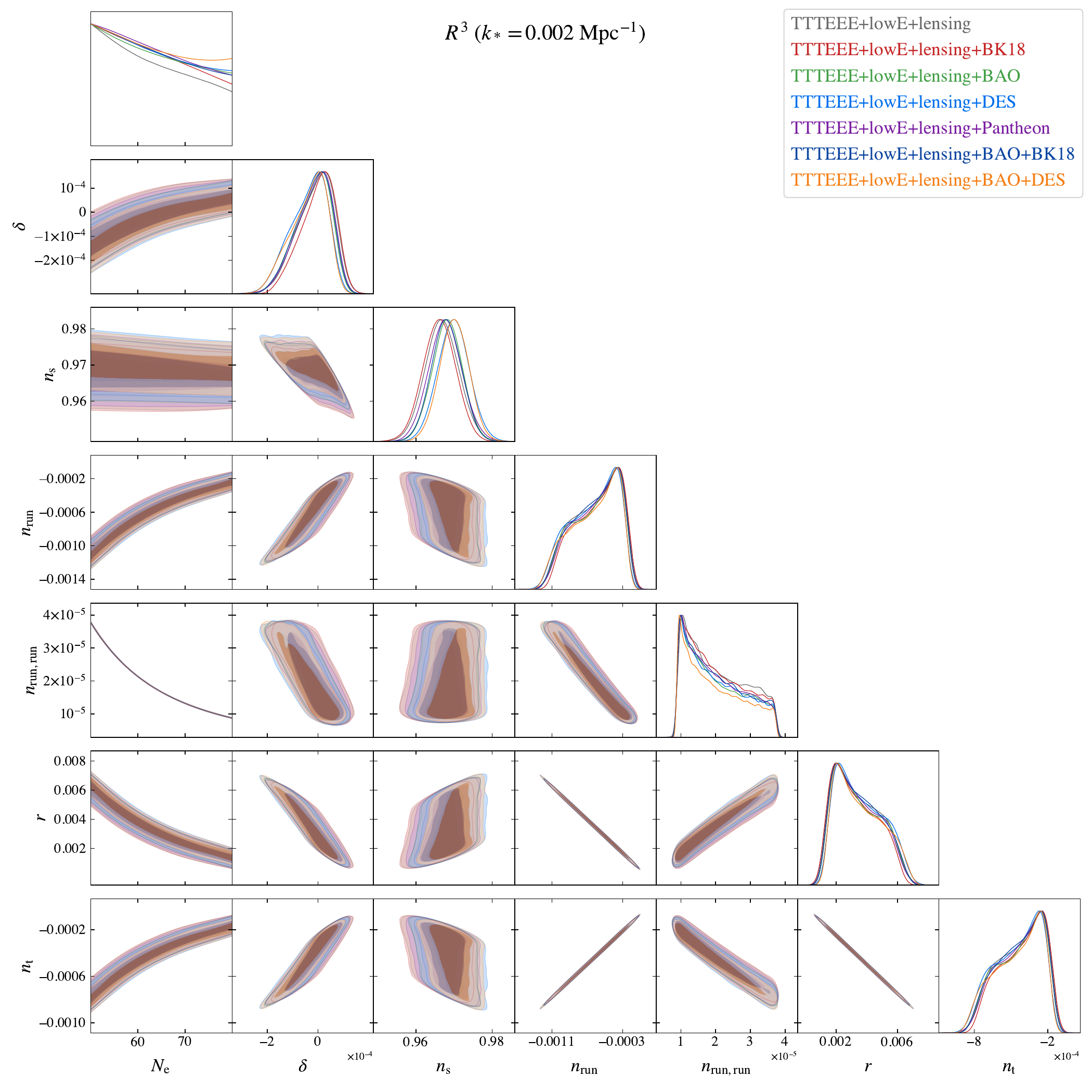}
\caption{\label{fig:all_data_exst_plot2} Same as \fref{fig:all_data_star_plot2} but for $R^3$ Starobinsky model.}
\end{figure*}

\newpage
${}$

\begin{figure*}
\includegraphics[width=\textwidth]{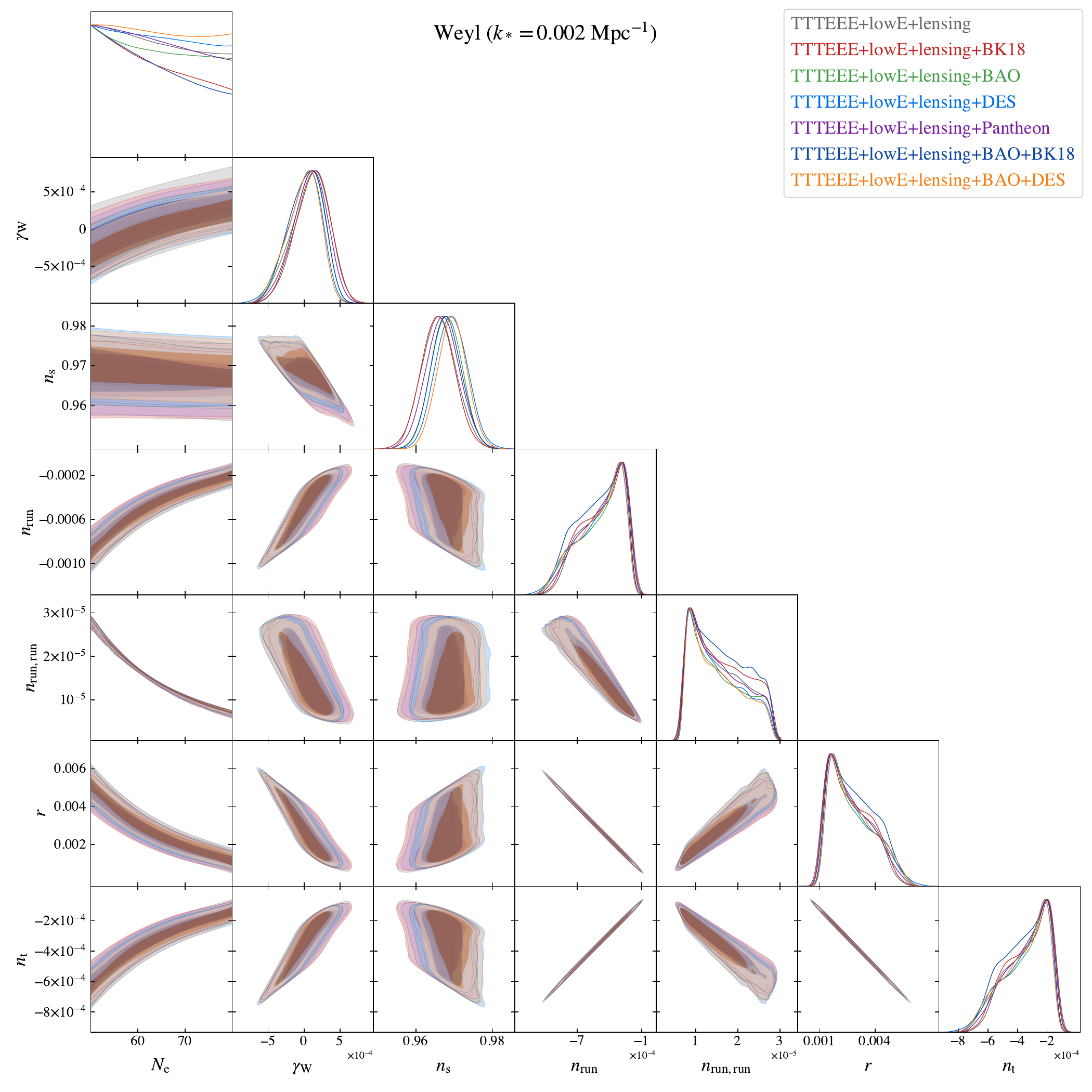}
\caption{\label{fig:all_data_weyl_plot2}
Same as \fref{fig:all_data_star_plot2} but for Weyl model.}
\end{figure*}

\end{document}